%% file: main.tex
\PassOptionsToPackage{table}{xcolor}
\documentclass[sigconf,screen]{acmart}
\usepackage{booktabs} 
\usepackage{rotating}
\usepackage{pdflscape}
\usepackage{fontawesome5}
\usepackage{lipsum}
\usepackage{enumitem}
\usepackage{textcomp}   
\usepackage{geometry}
\usepackage{xcolor}
\usepackage{colortbl}
\usepackage{caption}
\usepackage{ragged2e}
\usepackage[utf8]{inputenc}

\usepackage{array}

\definecolor{firsthand}{RGB}{33, 113, 181}    
\definecolor{secondhand}{RGB}{217, 95, 2}     
\definecolor{onchain}{RGB}{46, 204, 113}      
\definecolor{offchain}{RGB}{231, 76, 60}      
\definecolor{hybrid}{RGB}{241, 196, 15}       

\usepackage{tikz}
\usetikzlibrary{shapes.geometric, arrows.meta, positioning}
\usepackage{fontawesome5}

\tikzstyle{process} = [
  rectangle, 
  minimum width=2.8cm, 
  minimum height=1.1cm, 
  text centered, 
  draw=black, 
  font=\scriptsize,
  fill=gray!10
]
\tikzstyle{arrow} = [thick, ->, >=stealth]

\usepackage{longtable}
\usepackage{graphicx} 
\usepackage{amsmath}  

\usepackage{ragged2e}
\usepackage{multirow}
\usepackage{url}      
\usepackage{subcaption} 
\definecolor{gold}{RGB}{255,215,0}  
\definecolor{restrictive}{RGB}{255, 102, 102}  
\definecolor{moderate}{RGB}{255, 230, 128}    
\definecolor{supportive}{RGB}{153, 255, 153}  

\usepackage[utf8]{inputenc}  
\usepackage[greek,english]{babel}  

\AtBeginDocument{%
  }

\acmConference[Working paper]{Make sure to enter the correct
  conference title from your rights confirmation email}{July
  2025}{Metaverse}

\setcopyright{none}
\renewcommand\footnotetextcopyrightpermission[1]{}
\usepackage{hyperref}

\begin{document}

\title{SoK: Stablecoins for Digital Transformation — Design, Metrics, and Application with Real World Asset Tokenization as a Case Study}

\author[Luyao Zhang*]{Luyao Zhang}
\affiliation{%
  \department{Social Science Division and Digital Innovation Research Center}
  \institution{Duke Kunshan University}
  \country{China}
}

\authornote{Corresponds to: Luyao Zhang (email: lz183@duke.edu, address: Duke Kunshan University, No.8 Duke Ave. Kunshan, Jiangsu 215316, China.) }


\begin{abstract}
Stablecoins have emerged as a foundational element in the digital asset ecosystem, with their market capitalization surpassing \$230 billion as of May 2025. As fiat-referenced and programmable financial instruments, stablecoins offer a low-latency, globally interoperable infrastructure for payments, decentralized finance (DeFi), and tokenized commerce. Their rapid adoption has spurred significant regulatory attention, exemplified by the European Union’s Markets in Crypto-assets Regulation (MiCA), the U.S. Guiding and Establishing National Innovation for U.S. Stablecoins Act (GENIUS Act), and Hong Kong’s Stablecoins Bill. Despite this momentum, academic research on stablecoins remains siloed across economics, law, and computer science, lacking an integrated framework for design, evaluation, and application.

This study addresses this gap through a multi-method research design. First, it synthesizes cross-disciplinary literature to construct a multi-dimensional taxonomy of stablecoin systems, categorizing them by custodial structure, stabilization mechanism, and governance. Second, it introduces a performance metrics framework tailored to diverse stakeholder needs—regulators, developers, institutions, and end-users—supported by an open-source benchmarking pipeline to ensure transparency and reproducibility. Third, a case study on Real World Asset (RWA) tokenization illustrates how stablecoins serve as programmable monetary infrastructure in cross-border digital systems.

By integrating theoretical foundations with practical engineering tools, this paper contributes: (1) a unified taxonomy for stablecoin design; (2) a stakeholder-oriented performance evaluation framework; (3) an empirical case linking stablecoins to sectoral digital transformation; and (4) reproducible methodologies and datasets to guide future development. These contributions offer actionable insights for scaling stablecoins as trusted, transparent, and inclusive digital monetary infrastructure.
\end{abstract}





\maketitle

\section{Introduction}

Stablecoins have rapidly evolved into a critical component of the digital asset ecosystem. With a total market capitalization exceeding \$230 billion as of May 2025~\cite{coinmetrics2025}, stablecoins represent the largest class of crypto-assets used in digital payments, decentralized finance (DeFi), and tokenized commerce. Unlike volatile cryptocurrencies, stablecoins offer price stability, fiat-referenced value, and programmable interfaces, making them ideal monetary infrastructure for global transactions~\cite{catalini2022economics, klages2020stablecoins}.

This growing systemic importance has prompted global regulatory responses. In the European Union, the \textit{Regulation (EU) 2023/1114 on Markets in Crypto-assets (MiCA)} establishes a unified legal framework~\cite{mica2023}; in the United States, the \textit{Guiding and Establishing National Innovation for U.S. Stablecoins Act 2025 (GENIUS Act)} aims to guide and establish national innovation in stablecoin ecosystems~\cite{us_genius_s1582_2025}; while in Greater China, Hong Kong's Legislative Council passed the Stablecoins Bill (Bill) in May 2025~\cite{hkma2025}.

Despite this attention, the literature remains fragmented across disciplines. Studies in monetary economics~\cite{catalini2022economics, lyons2023what}, finance~\cite{ma2025stablecoin,harvey2024international}, and computer science~\cite{klages2020stablecoins} have advanced our understanding of stablecoin design, market behavior, and system vulnerabilities. Legal scholars highlight regulatory gaps and jurisdictional complexity~\cite{ferreira2021curious,schuler2024defi}. Recent empirical work also underscores the potential of DeFi and Web3 ecosystems—including stablecoins—to promote financial inclusion and democratized capital access, particularly in underserved regions~\cite{cong2023inclusion}. Yet, a comprehensive framework that connects stablecoin design to their enabling role in digital transformation is still missing. Additionally, existing research often lacks standardized, reproducible metrics and evaluation pipelines that could support both academic and industrial advancement.

This paper addresses these gaps by focusing on three core research questions:
\begin{enumerate}
    \item \textbf{Design:} What are the functional and architectural design patterns of stablecoins relevant for real-world digital transformation?
    \item \textbf{Metrics:} What core performance metrics can be developed to assess stablecoin systems from the perspectives of regulators, developers, users, and institutions?
    \item \textbf{Applications:} How can stablecoins be operationalized as digital monetary infrastructure in emerging sectors such as the low-altitude economy?
\end{enumerate}

To answer these questions, we employ a multi-method research design. First, in Section~\ref{sec:design}, we synthesize literature across finance, computer science, and legal studies to develop a multi-dimensional taxonomy of stablecoin systems, classifying them by custodial structure, stabilization mechanism, and governance~\cite{klages2020stablecoins, ferreira2021curious}. Next, in Section~\ref{sec: metrics}, we define and quantify a set of stakeholder-relevant performance metrics. These metrics are implemented using an open-source benchmarking pipeline with publicly available datasets to support reproducibility and comparative evaluation.

In Section~\ref{sec:applications}, we present a case study of stablecoin deployment in the context of \textit{Real World Asset (RWA) Tokenization}, demonstrating how stablecoins function as a trusted and programmable monetary layer within distributed financial infrastructure.

Our contributions are four-fold: (1) a unified taxonomy of stablecoin design, grounded in interdisciplinary synthesis and adaptable to evolving monetary architectures; (2) a stakeholder-oriented performance framework, supported by open-source implementation; (3) a sectoral case study that concretely links stablecoin infrastructure to digital transformation; and (4) a forward-looking discussion of design implications and research directions in Section~\ref{sec:future}.

All analyses presented in this work are fully reproducible. The complete dataset, preprocessing pipelines, and visualization scripts are openly accessible at \url{https://github.com/sunshineluyao/stablecoin}. This study aims to offer actionable insights for regulators, developers, and researchers working to responsibly scale stablecoin adoption for meaningful and inclusive socio-economic impact.

\section{Design}
\label{sec:design}

\subsection{From Traditional Currency to Scalable Digital Finance}

In classical monetary economics, money is defined by three core functions: a \textit{medium of exchange} (facilitating transactions), a \textit{store of value} (preserving purchasing power over time), and a \textit{unit of account} (providing a standard for pricing and recordkeeping) \cite{mishkin2007economics}. These fundamental needs have historically guided the design of sovereign \textbf{fiat currencies}—government-issued money not backed by physical commodities but by trust in the issuing authority.

To fulfill these traditional needs, fiat currencies were designed with three primary features:

\begin{itemize}
    \item \textbf{Stability} directly satisfies the need for a \textit{store of value} and \textit{unit of account}, by reducing volatility and enabling consistent pricing and wealth preservation.
    \item \textbf{Liquidity} fulfills the need for a \textit{medium of exchange}, allowing the currency to be quickly and efficiently transacted in the market.
    \item \textbf{Interest-yielding capacity} supports the \textit{store of value} function by incentivizing holding and investment through mechanisms like bank deposits or bonds.
\end{itemize}

While these design principles sufficed in traditional, jurisdiction-bound financial systems, evolving technological and societal conditions have introduced new, more complex requirements. As economies globalize and digitize, the financial system faces the need for \textbf{scalable finance}—systems that operate across borders, jurisdictions, and platforms \cite{schar2021decentralized}.

This transformation introduces three additional needs:

\begin{itemize}
    \item \textit{Global exchange}: the ability to conduct seamless, cross-border transactions without friction or reliance on national intermediaries.
    \item \textit{Automation intelligence}: the requirement that financial operations be capable of autonomous, programmable behavior through algorithms or smart contracts.
    \item \textit{Interoperability}: the need for financial assets and systems to integrate and function across diverse digital environments and protocols.
\end{itemize}

These new needs demand a rethinking of monetary instruments. Stablecoins—digitally-native assets typically pegged to fiat currencies—have emerged as a response \cite{bullmann2019stablecoins, gorton2021taming}. While early stablecoins mimicked fiat by providing \textbf{stability} and \textbf{liquidity} to meet traditional needs, the growing complexity of global finance now requires additional design features.

Modern stablecoins address both the classical and emerging needs through expanded properties:

\begin{itemize}
    \item \textbf{Global access} satisfies the need for \textit{global exchange}, enabling anyone with internet connectivity to participate in digital finance, regardless of geography or institutional affiliation \cite{li2021potential}.
    \item \textbf{Distributed trust} addresses both \textit{global exchange} and \textit{automation intelligence} by leveraging distributed ledger technologies to remove single points of control, enhance transparency, and enable trustless transactions \cite{werner2021sok}.
    \item \textbf{Programmability} directly meets the need for \textit{automation intelligence}, enabling assets to carry executable logic that supports conditional transfers, compliance rules, or financial contracts \cite{buterin2014ethereum}.
    \item \textbf{Composability} satisfies the need for \textit{interoperability} by allowing stablecoins to be integrated modularly into a variety of decentralized or platform-based financial systems \cite{werner2021sok}.
\end{itemize}

\begin{figure}
  \includegraphics[width=0.49\textwidth]{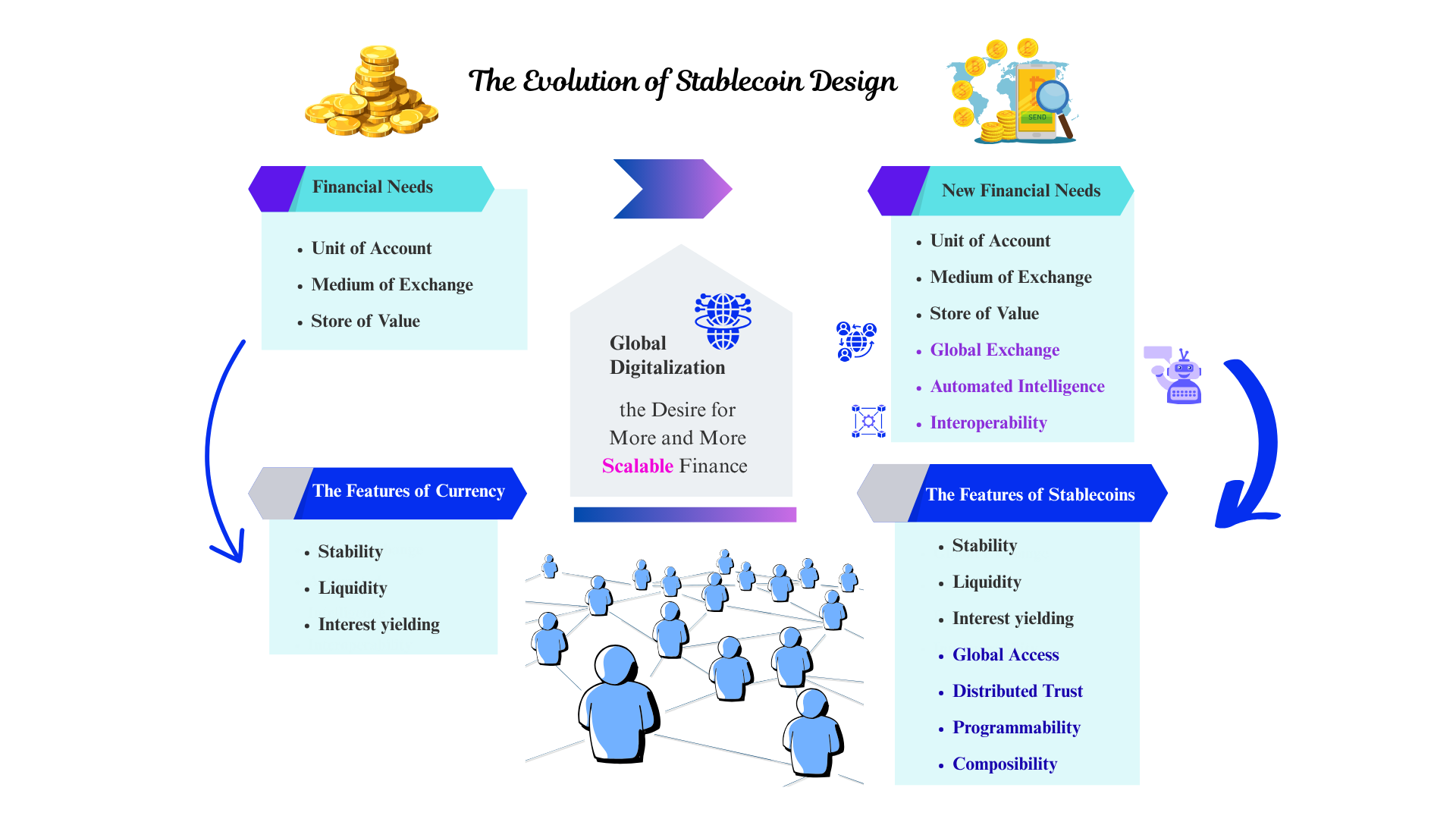}
  \caption{The Evolution of Stablecoin Design.}
  \label{fig:evolution}
\end{figure}

Figure~\ref{fig:evolution} visualizes this layered progression. The foundational needs of early economies drove the design of stable, liquid, interest-bearing currency. As global and digital demands intensified, stablecoins evolved by incorporating new design primitives to satisfy needs like \textit{global access}, \textit{distributed trust}, \textit{programmability}, and \textit{composability}. This reflects a broader evolutionary trajectory in financial systems, driven by human development toward global interconnectivity and machine-mediated economies.

\subsection{Stablecoin Definition and Objectives}

A \textit{stablecoin} is a blockchain-native digital asset designed to closely track the value of a reference unit—typically a fiat currency such as the U.S. dollar (USD) \cite{gorton2021taming, bullmann2019stablecoins}. Let $P_t$ denote the price of one unit of the stablecoin at time $t$, and let $P_t^{\text{peg}}$ represent the pegged fiat value (e.g., $P_t^{\text{peg}} = 1$ USD). The essential stability requirement is that $P_t$ remains close to $P_t^{\text{peg}}$ over time. This can be formalized as:

\begin{equation}
\mathbb{E}\left[ d(P_t, P_t^{\text{peg}}) \right] < \varepsilon,
\label{eq:general_deviation} 
\end{equation}

where $d(\cdot,\cdot)$ is a measure of deviation (e.g., absolute deviation or squared error), and $\varepsilon$ is a small threshold determined by system tolerance. 

\textit{Intuitively, this means that a stablecoin is successful if, on average, its market price stays very close to its intended peg, minimizing the user’s exposure to volatility.}

Stablecoins are transacted on both centralized exchanges (CEXs), which are subject to governmental regulations (e.g., Know Your Customer (KYC) and Anti-Money Laundering (AML) policies), and decentralized exchanges (DEXs), which operate via smart contracts on public blockchains. While CEX access entails regulatory restrictions, DEX access depends on users' technical proficiency, custody skills, and in-chain transaction fee management \cite{schar2021decentralized}. Stablecoins also leverage multi-blockchain architectures, a characteristic that introduces deployment complexity alongside enhanced composability.

USDC~\cite{circleUSDC}, a fiat-backed stablecoin, serves as an illustrative example of these operational characteristics. It is exchanged on numerous CEXs (e.g., Coinbase, Binance) and also functions on DEXs (e.g., Uniswap, Curve) for peer-to-peer trading and liquidity provision. Its deployment across multiple blockchains (e.g., Ethereum, Solana, Polygon, Avalanche) demonstrates stablecoins' role as a programmable monetary layer across various digital environments.

To fulfill both traditional monetary functions and modern digital finance demands, stablecoins must meet the following objectives:

\paragraph{1. Price Stability.}  
The fundamental design objective of a stablecoin is to minimize deviations from its predefined \textit{peg} value $P_t^{\text{peg}}$, which typically references a fiat currency (e.g., USD) or a commodity (e.g., gold). Importantly, a stablecoin’s peg is not necessarily equivalent to its reserve backing; for instance, \textbf{DAI} is crypto-backed via overcollateralized Ethereum-based assets, yet it pegs its value to the U.S. dollar~\cite{makerdaoWhitepaper}. Similarly, algorithmic or synthetic stablecoins may enforce a dollar peg without holding fiat reserves.

To formally evaluate price stability, we define the percentage deviation $\delta_t$ at time $t$ as:
\begin{equation}
\delta_t = \frac{P_t - P_t^{\text{peg}}}{P_t^{\text{peg}}} \times 100
\label{eq:percentage_deviation}
\end{equation}

Based on this definition, two widely adopted statistical measures are used to quantify price volatility around the peg:

\begin{itemize}
    \item \textbf{Mean Absolute Error (MAE)}:
    \begin{equation}
    \text{MAE} = \frac{1}{T} \sum_{t=1}^{T} |\delta_t|
    \end{equation}
    This metric captures the average absolute percentage deviation, offering a direct interpretation in relative terms. Lower values signify tighter peg maintenance.

    \item \textbf{Root Mean Squared Error (RMSE)}:
    \begin{equation}
    \text{RMSE} = \sqrt{\frac{1}{T} \sum_{t=1}^{T} \delta_t^2}
    \end{equation}
    RMSE penalizes larger deviations more heavily, emphasizing the severity of outliers in price fluctuation.
\end{itemize}

\paragraph{2. Liquidity.} Liquidity refers to the ease of converting stablecoins into other assets with minimal price impact. Let $P_t^{\text{ask}}$ and $P_t^{\text{bid}}$ be the lowest ask and highest bid prices in a given market at time $t$. Define the relative bid-ask spread as:

\begin{equation}
\delta_t = \frac{|P_t^{\text{ask}} - P_t^{\text{bid}}|}{P_t^{\text{peg}}}.
\label{eq:liquidity}
\end{equation}

\textit{A smaller spread $\delta_t$ means that users can buy and sell stablecoins at prices close to the peg, signaling healthy market depth and efficient price discovery.}

\paragraph{3. Yield Potential.} Some stablecoins provide a return on holdings, either through algorithmic interest mechanisms or through staking/lending. Let $r_t^{\text{stable}}$ be the stablecoin’s yield at time $t$, and $r_t^{\text{risk-free}}$ a benchmark rate (e.g., treasury bond yield). Then:

\begin{equation}
\mathbb{E}[r_t^{\text{stable}}] \geq r_t^{\text{risk-free}} + \eta,
\label{eq:interest}
\end{equation}

\textit{This ensures the stablecoin offers a competitive return, accounting for a small risk buffer $\eta$ due to smart contract or collateral risks.}

\paragraph{4. Global Accessibility.} Let $A(u, t, p) \in \{0,1\}$ indicate whether user $u$ can access the stablecoin on platform $p$ (either CEX or DEX) at time $t$. Then:

\begin{equation}
\Pr_{u \sim \mathcal{U}}\left[\exists p \in \{\text{CEX}, \text{DEX}\} : A(u,t,p) = 1 \right] \rightarrow 1,
\label{eq:access}
\end{equation}

\textit{This captures the goal that stablecoins should be globally accessible to internet-connected users, either through regulated exchanges or permissionless platforms.}

\paragraph{5. System-Enforced Trust.} Let $P_t$ denote the market price (peg value) of the stablecoin at time $t$, and $M_t$ represent the mechanism responsible for maintaining price stability. A stablecoin system is said to enforce trust if it ensures that $P_t$ remains consistently close to the intended peg:

\begin{equation}
\left| P_t - P_{\text{peg}} \right| \approx 0 \quad \text{under } M_t.
\label{eq:trust}
\end{equation}

\textit{That is, the system should preserve the redeemability of the stablecoin at or near its target value through verifiable means. These mechanisms may be collateral-based (e.g., fiat, crypto, or tokenized real-world assets), algorithmic (e.g., mint-and-burn or supply contraction/expansion rules), or hybrid architectures. In collateralized systems, the \textbf{collateral ratio}—defined as the ratio of backing asset value to circulating stablecoins—can serve as a key metric of trustworthiness. Mechanisms may be enforced or supported by centralized agents (such as custodians or auditors) and/or decentralized protocols (such as oracles, governance modules, and smart contracts). Robust systems minimize single points of failure while ensuring redemption confidence across operational states~\cite{werner2021sok}.}

\paragraph{6. Programmability.} Let $f: \mathcal{E} \rightarrow \mathcal{A}$ be a deterministic mapping from on-chain events $\mathcal{E}$ to actions $\mathcal{A}$. Then:

\begin{equation}
f(e) \text{ executes automatically and verifiably on-chain, } \forall e \in \mathcal{E}.
\label{eq:programmability}
\end{equation}

\textit{This capability allows stablecoins to participate in smart contract logic—such as conditional payments, lending automation, or compliance enforcement \cite{buterin2014ethereum}.}

\paragraph{7. Composability.} Let $\mathcal{P}$ be a set of DeFi protocols. Define $C(P_t, \pi_j) \in \{0,1\}$ as an indicator of whether the stablecoin is supported in protocol $\pi_j$. Then:

\begin{equation}
\forall \pi_j \in \mathcal{P}, \quad C(P_t, \pi_j) = 1.
\label{eq:composability}
\end{equation}

\textit{Composability ensures the stablecoin integrates seamlessly across the DeFi stack—serving as collateral, medium of exchange, or liquidity asset without barriers \cite{werner2021sok}.}

\bigskip

Together, these properties characterize the multi-dimensional design space of modern stablecoins. Stability, liquidity, and yield ensure economic viability; global access and distributed trust promote inclusivity and resilience; while programmability and composability ensure integration into the broader programmable financial ecosystem.

\subsection{Stablecoin Design Taxonomy}

To implement the system objectives described in the previous section—including \textit{price stability}, \textit{liquidity}, \textit{yield potential}, \textit{global accessibility}, \textit{distributed trust}, \textit{programmability}, and \textit{composability}—stablecoins employ a combination of economic, technical, and governance mechanisms. We categorize these mechanisms into key \textit{design facets}, each of which addresses one or more objectives.

The taxonomy in Table~\ref{tab:taxonomy} presents these design choices, aligning them with representative examples. Below, we elaborate on each facet, referencing the formal objectives defined in Section~\ref{sec:design}.

\subsubsection{Stabilization Mechanism}

The \textit{stabilization mechanism} defines how a stablecoin maintains its target price (e.g., \$1.00 or 1 ounce of gold), which directly affects \textbf{price stability} as captured by Equation~\ref{eq:general_deviation}.

\begin{itemize}
    \item \textbf{Fiat-backed:} Collateralized 1:1 with fiat currency reserves held in bank accounts. \textbf{USD Coin (USDC)}~\cite{circleUSDC} pegs to the U.S. dollar through regulated custodians.
    \item \textbf{Crypto-backed:} Overcollateralized with crypto assets on-chain. \textbf{DAI}~\cite{makerdaoWhitepaper} uses ETH and USDC as collateral.
    \item \textbf{Algorithmic (+ crypto-backed):} Partial reserves combined with algorithmic supply control. \textbf{FRAX (FRAX)}~\cite{fraxDocs} integrates AMOs and governance.
    \item \textbf{Synthetic:} Pegs are maintained via market hedging (e.g., short perpetual futures). \textbf{USDe}~\cite{ethenaDocs} uses ETH and delta-neutral strategies.
    \item \textbf{Commodity-backed:} Pegged to physical commodities like gold or silver. Examples include \textbf{PAX Gold (PAXG)}~\cite{paxgWhitepaper}, backed 1:1 by LBMA-approved gold held in Brink’s vaults; \textbf{Tether Gold (XAUT)}~\cite{xautWhitepaper}, redeemable for physical gold; and \textbf{Kinesis Silver (KAG)}~\cite{kinesisDocs}, pegged to physical silver reserves.
\end{itemize}

\subsubsection{Custodial Structure}

The \textit{custodial structure} defines how reserve assets are held and who controls them. This is central to \textbf{distributed trust}, where the influence of any single agent $T_i$ is constrained as defined by Equation~\ref{eq:trust}.

\begin{itemize}
    \item \textbf{Centralized:} Managed by institutions under legal agreements. \textbf{USDC}~\cite{circleUSDC} is custodied by BNY Mellon and BlackRock.
    \item \textbf{On-chain collateral:} Controlled transparently by smart contracts. \textbf{DAI}~\cite{makerdaoWhitepaper} supports full on-chain visibility and access control.
    \item \textbf{Algorithmic reserves:} Enforced via protocol-defined hedging strategies. \textbf{USDe}~\cite{ethenaDocs} uses smart contracts without central custodians.
    \item \textbf{Hybrid:} Combines centralized management with on-chain components. \textbf{FRAX}~\cite{fraxDocs} uses both Frax DAO governance and smart contracts.
\end{itemize}

\subsubsection{Interest Mechanism}

The \textit{interest mechanism} governs whether yield from reserves is shared with users, supporting \textbf{yield potential} as specified in Equation~\ref{eq:interest}.

\begin{itemize}
    \item \textbf{No yield:} Yield retained by issuer. \textbf{Classic USDC}~\cite{circleUSDC} does not distribute earnings from its Treasury holdings.
    \item \textbf{RWA-based yield:} Interest from real-world assets (RWAs) like Treasury bills is shared with users. \textbf{Institutional USDC} \cite{circleUSDC} is designed for this.
    \item \textbf{Crypto-collateralized yield:} Derived from lending or staking rewards. \textbf{sDAI}~\cite{makerdaoWhitepaper} pays interest via MakerDAO's DAI Savings Rate.
    \item \textbf{Synthetic yield:} Generated from derivatives-based strategies. \textbf{sUSDe}~\cite{ethenaDocs} offers yield backed by synthetic arbitrage and leverage.
\end{itemize}

\subsubsection{Market Access}

The \textit{market access} dimension refers to how users interact with the stablecoin, contributing to \textbf{global accessibility} as defined in Equation~\ref{eq:access}.

\begin{itemize}
    \item \textbf{Single-region CEX listing:} Limited to a small number of exchanges in specific jurisdictions. \textbf{PayPal USD (PYUSD)}~\cite{paypal2023pyusd} is primarily listed on U.S. platforms.
    \item \textbf{Multi-CEX global listing:} Widely available across centralized exchanges. \textbf{Tether (USDT)}~\cite{tetherWhitepaper} is accessible on platforms like Binance, OKX, and Kraken.
\end{itemize}

\subsubsection{Governance Model}

The \textit{governance model} defines how protocol decisions are made, influencing both \textbf{distributed trust} and \textbf{programmability} through the execution logic $f: \mathcal{E} \rightarrow \mathcal{A}$ as expressed in Equation~\ref{eq:programmability}.

\begin{itemize}
    \item \textbf{Corporate:} Governance is executed by a private company. \textbf{USDC}~\cite{circleUSDC} is controlled by Circle.
    \item \textbf{DAO:} Decisions are voted on by token holders. \textbf{DAI}~\cite{makerdaoWhitepaper} is governed by MakerDAO via MKR.
    \item \textbf{Hybrid:} Combines DAO decision-making with executive authority. \textbf{FRAX}~\cite{fraxDocs} uses both Frax DAO and off-chain management.
\end{itemize}

\subsubsection{Protocol Interoperability}

\textit{Protocol interoperability} refers to how well a stablecoin integrates into different DeFi systems, fulfilling the objective of \textbf{composability}, defined by Equation~\ref{eq:composability}.

\begin{itemize}
    \item \textbf{Single-chain integration:} Operates on a single blockchain. \textbf{PYUSD}~\cite{paypal2023pyusd} is deployed solely on Ethereum.
    \item \textbf{Multi-chain integration:} Issued on several blockchains. \textbf{USDC}~\cite{circleUSDC} is native to Ethereum, Solana, Avalanche, Arbitrum, and others.
\end{itemize}

\subsubsection{Liquidity Mechanism}

The \textit{liquidity mechanism} governs how stablecoins are traded and how slippage is minimized, supporting the objective of \textbf{liquidity}, as defined in Equation~\ref{eq:liquidity}.

\begin{itemize}
    \item \textbf{CEX double auction:} Executed via centralized order books. \textbf{USDT}~\cite{tetherWhitepaper} is deeply liquid on CEXs like Binance and OKX.
    \item \textbf{AMM (DEX pool):} Prices determined by on-chain liquidity curves. \textbf{DAI}~\cite{makerdaoWhitepaper} is liquid on DEXs like Uniswap and Curve.
\end{itemize}

\bigskip

Taken together, these facets define a modular blueprint for stablecoin design. Each mechanism contributes to one or more objectives outlined in Section~\ref{sec:design}, and the concrete examples in Table~\ref{tab:taxonomy} illustrate the trade-offs and innovations shaping modern programmable money.

\section{Metrics}
\label{sec: metrics}

To enable reproducible and comparative evaluation of stablecoin systems, we define four core indices rooted in four fundamental design goals: \textit{price stability}, \textit{distributed trust}, \textit{yield potential}, and \textit{market accessibility}. These indices are scoped to support general comparability across stablecoin projects. We explicitly defer infrastructure-specific, microstructural indices to future work (see Section~\ref{sec:future}).

\subsection{Data Provenance and Classification}

We categorize all data used for metric construction based on its origin and method of aggregation:

\begin{itemize}
  \item \textbf{\textcolor{onchain}{\faCube} On-chain:} Native blockchain data queried via RPC interfaces such as Infura~\cite{infuraDocs} and Alchemy~\cite{alchemyDocs}, or via libraries like Web3.js~\cite{web3jsDocs}, Graph~\cite{thegraphDocs} and ethers.js~\cite{ethersDocs}. These tools allow contract interaction and event log extraction. 

  \item \textbf{\textcolor{offchain}{\faCircle} Off-chain:} Centralized APIs and disclosures (e.g., SEC filings~\cite{secEDGAR}) that provide institutional user base estimates and exchange metadata.

  \item \textbf{\textcolor{hybrid}{\faAdjust} Hybrid:} Curated analytics services such as CoinGecko \cite{coingeckoAPI}, DeFiLlama~\cite{defillama2024}, Chainlink \cite{chainlinkDocs}, CoinMetrics~\cite{coinmetrics2024}, Bitquery~\cite{bitqueryDocs}, Google BigQuery~\cite{googleBigQuery}, Dune Analytics~\cite{duneAnalytics}, and AWS Open Data Registry~\cite{awsOpenData}.
\end{itemize}

Table~\ref{tab:stablecoin_data_sources} provides a detailed summary.

\subsection{Metric Definitions}

Each metric is defined over a measurement window \(T\), using data resampled to a minimum temporal resolution \(\tau_{\min}\). This ensures consistency across protocols that differ in data availability and sampling granularity.

\subsubsection{Price Stability (RMSE)}  
To quantitatively assess how closely a stablecoin maintains its target peg, we define the pointwise deviation at time \(t\) as:
\begin{equation}
\delta_t = \frac{P_t - P^{\text{peg}}}{P^{\text{peg}}}
\label{eq:price_deviation}
\end{equation}
where \(P_t\) is the observed market price at time \(t\) and \(P^{\text{peg}}\) denotes the nominal peg value (e.g., \$1.00 for most fiat-pegged coins). The root-mean-square error (RMSE) over the evaluation window \(T\) is then:
\begin{equation}
\text{RMSE}_T = \sqrt{\frac{1}{|T|} \sum_{t \in T} \delta_t^2}
\label{eq:rmse}
\end{equation}
This metric reflects both the magnitude and frequency of deviation from the peg.

It is critical to distinguish between a stablecoin's \textit{peg} and its \textit{collateral backing}. While most stablecoins—including fiat-backed (e.g., USDC, USDT, BUSD, GUSD, PAX), crypto-backed (e.g., DAI), and commodity-backed (e.g., PAXG, XAUT) types—seek to maintain a stable reference value, the nature of their reserve assets and operational risks differ significantly. For instance, DAI is overcollateralized using crypto assets such as ETH and USDC~\cite{makerdaoWhitepaper}, whereas USDC and GUSD are fully backed by fiat reserves held in regulated financial institutions. PAXG~\cite{paxgDocs} and XAUT~\cite{xautDocs} are tokenized representations of gold, pegged to the daily market price of one troy ounce, and are subject to commodity market fluctuations rather than monetary policy anchoring.

Table~\ref{tab:peg_deviations} presents a comparative snapshot of select fiat-pegged and commodity-pegged digital assets as of 2025-07-15. The table includes:
\begin{itemize}
  \item The \textbf{design type} (e.g., fiat-backed, crypto-backed, gold-backed),
  \item The \textbf{peg reference} (USD or Gold),
  \item The nominal \textbf{peg value} in USD,
  \item The \textbf{observed market price}, retrieved from CoinMetrics' community API via the \texttt{PriceUSD} VWAP metric,
  \item The computed \textbf{deviation percentage}, based on Equation~\ref{eq:price_deviation}.
\end{itemize}

USD-based stablecoins are evaluated against a fixed peg of \$1.00. Gold-backed tokens are evaluated against a derived peg based on the daily spot price of gold. This was calculated using data from the National Bank of Poland (NBP)~\cite{nbpGoldAPI, nbpFXAPI}, converting PLN per gram rates to USD per troy ounce using:
\[
\text{USD/oz} = (\text{PLN/g}) \times (\text{USD/PLN}) \times 31.1035
\]

VWAP (Volume Weighted Average Price) was used as the pricing metric~\cite{coinmetrics_vwap}, ensuring that high-volume trades exert proportionally greater influence on the reported price—providing a more representative measure of market value than last-trade prices or midpoints.

In this empirical snapshot, major fiat-pegged stablecoins—including USDC, DAI, USDT, GUSD, and BUSD—exhibited very small deviations from their pegs (\(< \pm0.12\%\)), highlighting their robustness in maintaining price parity under prevailing market conditions. Notably, PAX deviated by approximately \(-0.23\%\), suggesting mild slippage likely due to liquidity factors. In contrast, gold-backed tokens (PAXG, XAUT) showed larger positive deviations (+1.24\% and +0.82\%, respectively), attributable to intraday gold price volatility, custody costs, and reduced secondary market liquidity.

\subsubsection{Distributed Trust via Collateral Ratios}
\label{sec:collateral_definitions}

A collateral ratio measures the degree to which a stablecoin is backed by verifiable assets. Formally, it is defined as:
\begin{equation}
\text{CR}_T = \frac{C_T}{S_T}
\label{eq:cr}
\end{equation}
where $C_T$ represents the total value of reserves held at time $T$, and $S_T$ denotes the market capitalization (supply) of the stablecoin at the same time. A collateral ratio of 1.00 implies full backing, while a ratio above 1.00 reflects overcollateralization.

This metric is particularly significant for fiat-backed or commodity-backed stablecoins, which rely on off-chain custodianship and third-party attestations. Because their collateral exists outside of a trustless blockchain environment, users must rely on external institutions—including financial regulators and accounting firms—to validate that sufficient, liquid, and appropriately risk-managed reserves exist. While not the only avenue for assessing distributed trust (others include decentralized governance, code audits, and price oracles), the collateral ratio offers a quantitative foundation for evaluating solvency risk in fiat-collateralized models.

However, the ratio itself must be interpreted in light of reserve quality, auditor credibility, and jurisdictional oversight. For instance, a 1.05 collateral ratio consisting of illiquid assets such as commercial paper offers less protection than a 1.00 ratio composed entirely of short-term U.S. Treasuries. Thus, the collateral ratio provides necessary—but not sufficient—evidence of sound financial design.

\subsubsection{Yield Potential (DeFi APY)}  
Unlike traditional fiat currencies stored in bank accounts, stablecoins such as USDC can be deployed in decentralized finance (DeFi) ecosystems to generate on-chain yield. This yield does not arise from centralized monetary policy or banking intermediation, but instead from a variety of protocol-level mechanisms—including peer-to-peer lending, automated market-making, liquidity mining, and synthetic asset exposure. The result is a system where stablecoins serve not only as a digital store of value but also as a native yield-bearing instrument within blockchain economies.

\paragraph{Mathematical Formulation.}
Let $P$ denote the principal amount of USDC deposited into a DeFi protocol, and let $r(t)$ represent the instantaneous yield rate at time $t$. Assuming continuous compounding, the future value $V(t)$ of a deposit evolves as:
\[
V(t) = P \cdot e^{\int_0^t r(s)\, ds}
\]
In most practical scenarios, yield is reported in discrete terms using the Annual Percentage Yield (APY). For a pool $i$ under protocol $j$ deployed on blockchain $k$, the expected one-year return is modeled as:
\[
V_{ijk}(1\,\text{yr}) = P_{ijk} \cdot (1 + r_{ijk})
\]
where $r_{ijk}$ is the observed APY, typically calculated from block-by-block reward distributions, lending rates, or performance fees visible on-chain.

\paragraph{DeFi Yield Architecture.}
Each DeFi protocol comprises one or more pools into which users may deposit stablecoins like USDC. These protocols are categorized by their functional design—e.g., lending, yield optimization, or real-world asset (RWA) financing—and are deployed across different blockchains. Each layer of this structure influences both the total value locked (TVL) and the yield potential of user capital.

\begin{figure*}[!t]
    \centering
    \includegraphics[width=\textwidth]{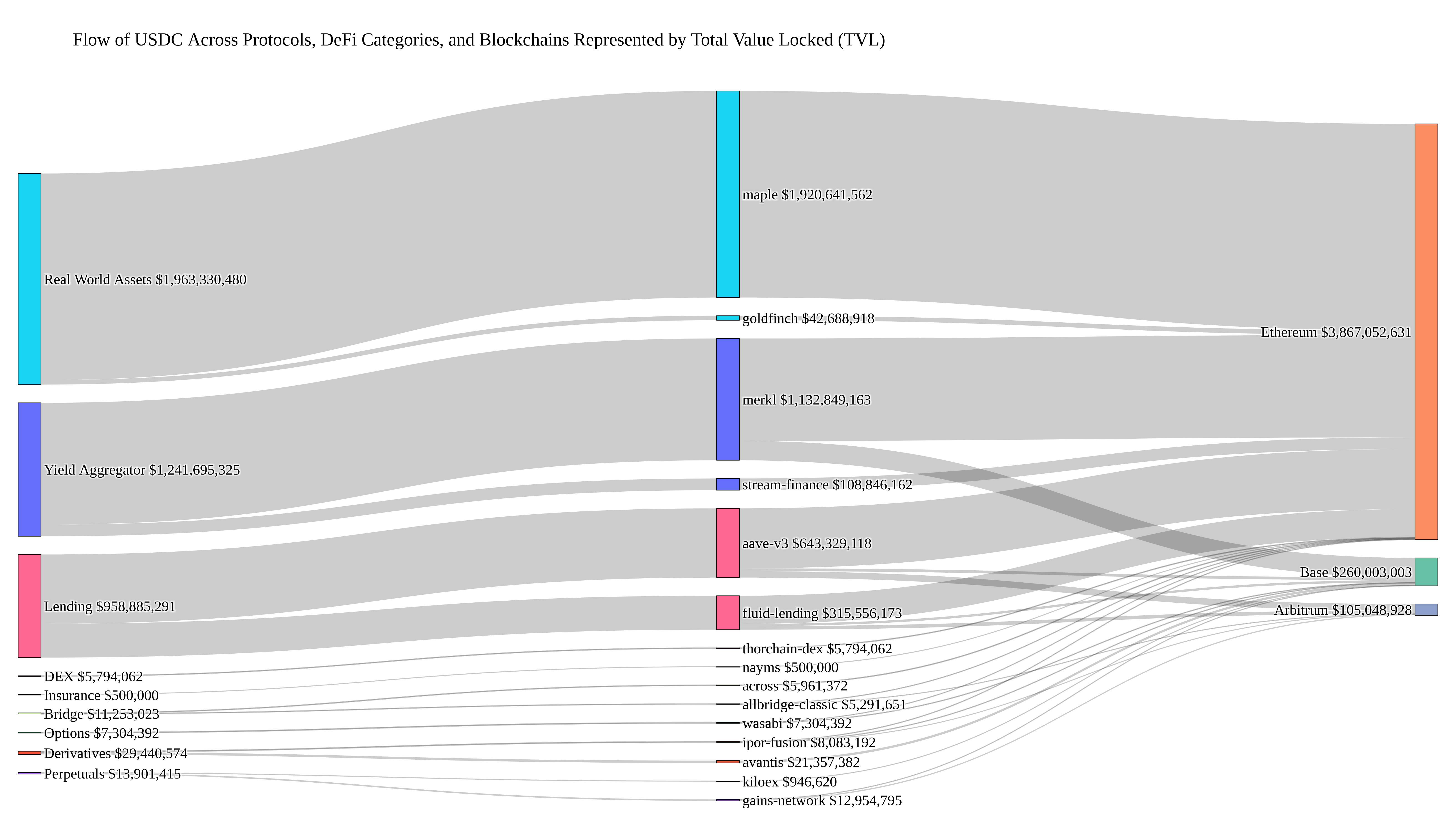}
    \caption{USDC Total Value Locked (TVL) Flow showing the relationship from protocol to category to blockchain. For clarity, the top two protocols by TVL per category are shown, along with the top three blockchain platforms where these protocols are deployed. Each node is labeled with the aggregated TVL in USDC, and color represents DeFi category. Data reflects the state of on-chain pools from \texttt{yields.llama.fi/pools} as of August 3, 2025.}
    \label{fig:usdc-tvl-sankey}
\end{figure*}

The empirical data analyzed in this study was obtained from the open API at \texttt{yields.llama.fi/pools}, and queried on \textbf{August 3, 2025}. This dataset provides granular, on-chain information for active decentralized finance (DeFi) pools, including protocol-level and pool-level metrics such as annual percentage yield (APY), total value locked (TVL), token denomination, and chain deployment.

Figure~\ref{fig:usdc-tvl-sankey} presents a high-resolution Sankey diagram visualizing the capital allocation of USDC across different layers of the DeFi stack. For clarity and interpretability, the diagram includes only the top two protocols by USDC TVL from each major DeFi category, and restricts blockchain platforms to the top three by cumulative TVL. Notably, in the Lending category, \textit{Aave V3} and \textit{Fluid Lending} account for over $950$ million in combined deposits. In the Yield Aggregator category, protocols such as \textit{Merkl} and \textit{Stream Finance} manage over $1.2$ billion in USDC. The Real World Assets (RWA) segment is represented by \textit{Maple} and \textit{Goldfinch}, with Maple alone holding nearly $2$ billion in deposits. Flow widths are proportional to aggregated TVL, and node labels include dollar-denominated values for each entity. The categorical distinctions are visually encoded through color, and terminal nodes on the right represent deployment blockchains—namely Ethereum, Base, and Arbitrum.

\paragraph{Heterogeneity in Yield Generation.}
While TVL represents the magnitude of capital commitment within a protocol, the APY quantifies its yield-generating effectiveness. These two variables do not always correlate, as they depend on protocol mechanics, pool utilization, risk exposure, and market demand for credit or liquidity.

\begin{figure}[!t]
\centering
\includegraphics[width=0.48\textwidth]{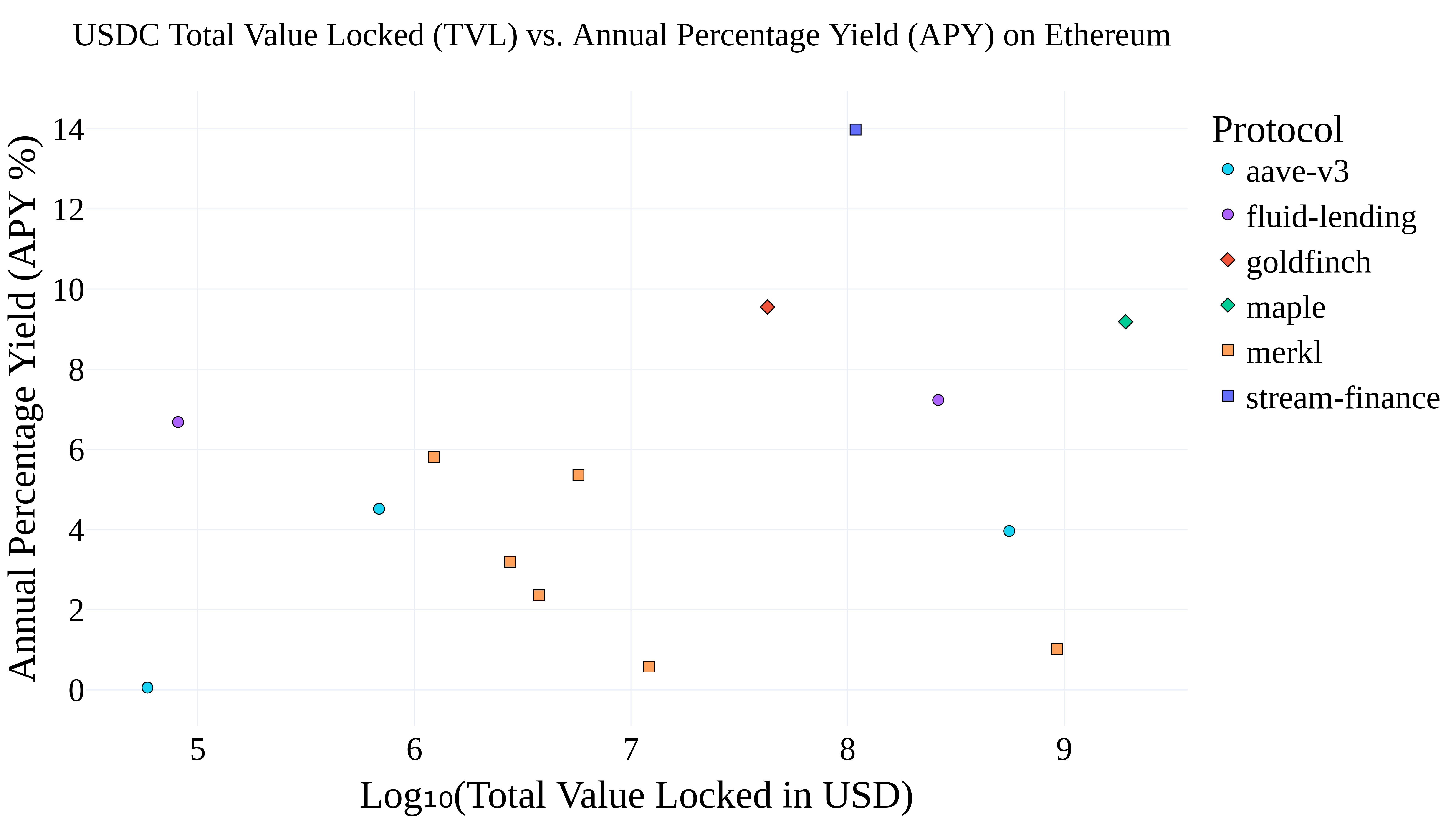}
\caption{USDC APY versus TVL across selected DeFi protocols on Ethereum. Marker color represents the protocol; marker shape denotes the DeFi category: circle (Lending), square (Yield Aggregator), and diamond (Real World Assets). Data reflects the state of on-chain pools from \texttt{yields.llama.fi/pools} as of August 3, 2025.}
\label{fig:eth-defi-apy-tvl}
\end{figure}

To examine this dynamic more closely, Figure~\ref{fig:eth-defi-apy-tvl} plots APY against the logarithm of TVL for selected USDC pools deployed on the Ethereum blockchain. The figure focuses on six protocols across three DeFi categories: Lending (\textit{Aave V3}, \textit{Fluid Lending}), Yield Aggregators (\textit{Merkl}, \textit{Stream Finance}), and Real World Assets (\textit{Maple}, \textit{Goldfinch}). Each protocol is represented by a distinct color, while marker shapes denote the DeFi category—circles for Lending, squares for Yield Aggregators, and diamonds for RWA.

For example, \textit{Aave V3} offers a broad pool structure with multiple deployment options and yields in the 2–6\% range, while \textit{Fluid Lending} operates a more concentrated deployment with moderate APYs. In the RWA category, both \textit{Maple} and \textit{Goldfinch} maintain a single pool each, offering relatively higher APYs (~9–14\%) commensurate with their off-chain credit underwriting risks. Yield Aggregators such as \textit{Merkl} display wider dispersion in APY due to dynamic strategy allocation and incentive programs. The use of log-transformed TVL on the x-axis aids interpretability across orders of magnitude in capital commitment.

Together, these two visualizations illustrate the layered architecture of yield generation in decentralized finance. Yield outcomes are not solely dictated by protocol design, but emerge from an interdependent structure involving category-level mechanisms, capital concentration, and the infrastructure of the hosting blockchain—offering a transparent, composable, and programmable alternative to traditional financial systems.

\subsection{Computation Workflow}

All metrics are computed using the pipeline in Table~\ref{tab:metric_pipeline}. 

\subsubsection{Price Stability}

To evaluate the price stability of USD-pegged stablecoins, we compute the deviation of each token’s observed market price \(P_t\) from its nominal peg value \(P^{\text{peg}} = \$1.00\), using the deviation formula in Equation~\ref{eq:price_deviation}. Deviations are calculated at daily intervals over a three-month observation window ending on 2025-08-01, based on volume-weighted average price (VWAP) data sourced from the CoinMetrics API.

The root-mean-square error (RMSE) of these deviations—defined in Equation~\ref{eq:rmse}—is used to summarize overall peg performance. RMSE jointly captures both the magnitude and persistence of deviations from the target price.

\begin{figure}[htbp]
\centering
\includegraphics[width=\linewidth]{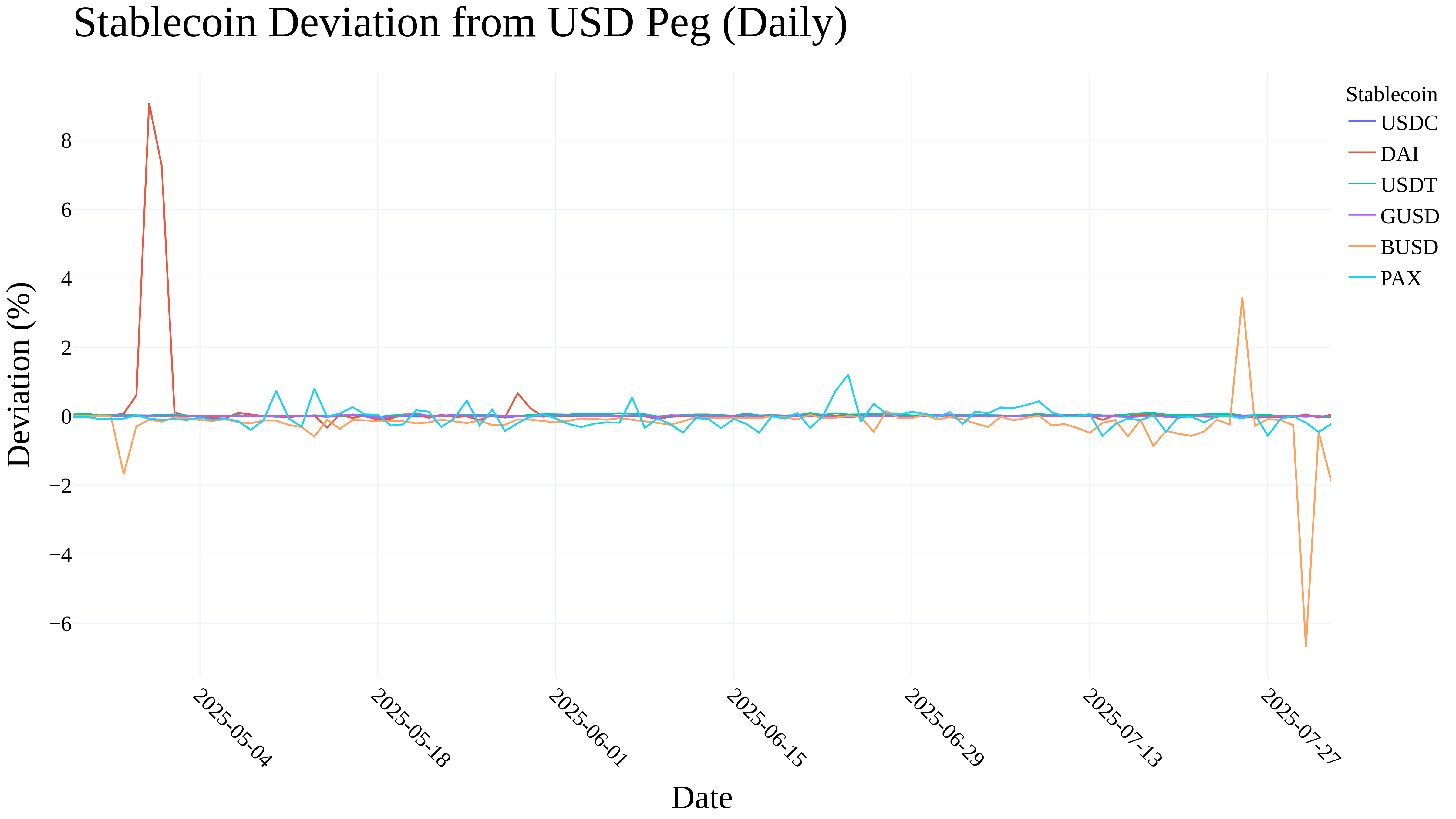}
\caption{Daily deviation from USD peg (\%) for selected stablecoins between May 1, 2025 and August 1, 2025. Most tokens fluctuate within a tight deviation band (\(< \pm 0.05\%\)), although some (e.g., BUSD, DAI) exhibit elevated volatility. Data Sources: CoinMetrics API.}
\label{fig:deviation_timeseries}
\end{figure}

\begin{figure}[htbp]
\centering
\includegraphics[width=\linewidth]{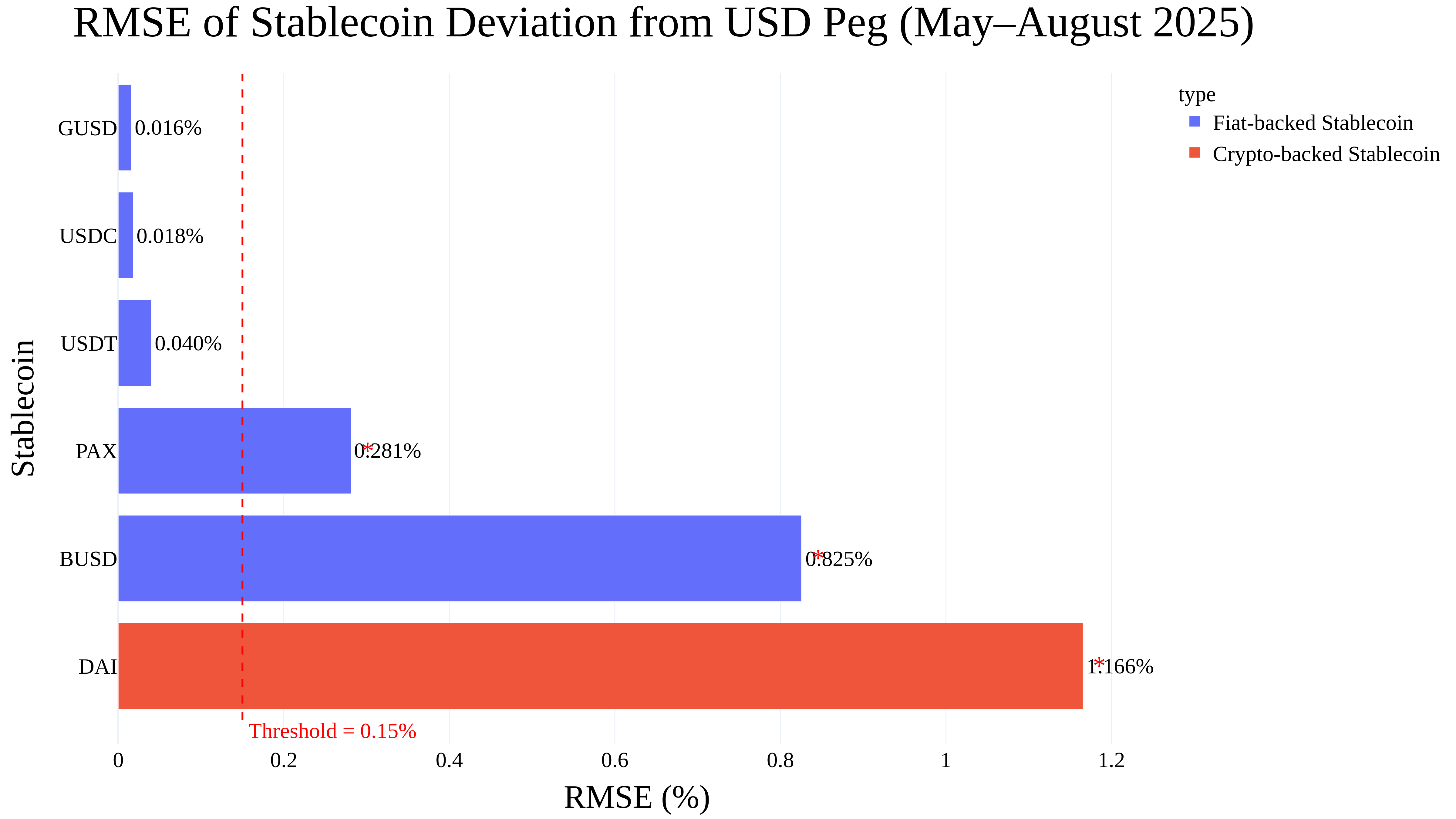}
\caption{Root-mean-square error (RMSE) of price deviation from USD peg for selected stablecoins (May–August 2025). A red dashed vertical line at 0.15\% indicates a tolerance threshold. Tokens with RMSEs above this line are marked with asterisks (*). Data Sources: CoinMetrics API.}
\label{fig:rmse_summary}
\end{figure}

Figure~\ref{fig:deviation_timeseries} illustrates the day-to-day deviation percentages, and Figure~\ref{fig:rmse_summary} ranks the tokens by RMSE, highlighting those exceeding a 0.15\% tolerance threshold.

Among fiat-backed tokens, GUSD (\(0.0158\%\)) and USDC (\(0.0179\%\)) demonstrate the highest peg fidelity, maintaining price stability within two basis points over the full period. Both are backed by regulated fiat reserves—GUSD via Gemini Trust and USDC via Circle—and are supported on major centralized exchanges such as Coinbase, Kraken, and Binance, and across multiple blockchains (e.g., Ethereum, Solana, Avalanche).

USDT, the most traded stablecoin globally, shows slightly greater volatility (RMSE \(= 0.0400\%\)), though still well within the tolerance band. Despite questions over its reserve transparency in the past, its liquidity dominance likely helps anchor its peg through market efficiency.

In contrast, legacy assets like PAX (\(0.2811\%\)) and especially BUSD (\(0.8255\%\)) show substantial peg drift. Both are issued by Paxos and fully fiat-collateralized, but BUSD's decline is likely driven by halted new issuance and delistings from major platforms, which reduced liquidity and arbitrage incentives.

DAI, the only crypto-backed token in the sample, has the highest RMSE at \(1.1657\%\). While it uses overcollateralized assets such as ETH and USDC, its reliance on decentralized governance, fluctuating collateral valuations, and indirect exposure to fiat markets may introduce instability during market dislocations.

Overall, the results suggest that fiat-backed stablecoins with active exchange support and multichain presence exhibit superior peg stability. Conversely, regulatory intervention, shrinking liquidity, or more complex reserve structures—as in the case of DAI—may lead to more pronounced peg deviations over time.

\subsubsection{Collateral Ratio}

Table~\ref{tab:collateral_comparison} summarizes the reported collateralization data across major fiat-backed stablecoins, including reserve composition, issuer identity, and the attesting auditor or regulatory body. To evaluate distributed trust through the lens of collateralization, we collected transparency data from the latest available reports on major fiat-backed stablecoins, including USDC, USDT, GUSD, BUSD, and USDP. Each report specifies the date of attestation, the reported reserves in U.S. dollars, and the circulating supply, from which we computed the collateral ratio using Equation~\ref{eq:cr}.

All stablecoins analyzed report full collateralization, with ratios ranging from 100.00\% to 103.50\%. USDC maintains a modest overcollateralization at 100.47\% (report dated July 31, 2025~\cite{circle2025}), backed primarily by U.S. Treasuries and overnight repurchase agreements managed within the Circle Reserve Fund, which is regulated by the U.S. Securities and Exchange Commission (SEC) and operated by BlackRock. USDT, the most capitalized stablecoin, reports a collateral ratio of 103.50\% (June 30, 2025~\cite{tether2025}), with a more diversified reserve structure including T-bills, cash deposits, gold, and Bitcoin. These reserves are attested to by BDO Italia.

GUSD, issued by Gemini Trust Company, LLC, maintains a strict 1:1 backing with fiat equivalents and is attested by BPM LLP, a certified public accounting firm based in the United States (May 30, 2025~\cite{gusd2025}). Paxos-issued BUSD and USDP both maintain full backing with cash and equivalents, and their reports are verified by WithumSmith+Brown PC. BUSD's report was dated June 30, 2025~\cite{busd2025}, and USDP's was also dated June 30, 2025~\cite{usdp2025}. The collateral ratios for BUSD and USDP stand at 100.35\% and 100.00\%, respectively.

\begin{figure}[htbp]
    \centering
    \includegraphics[width=0.48\textwidth]{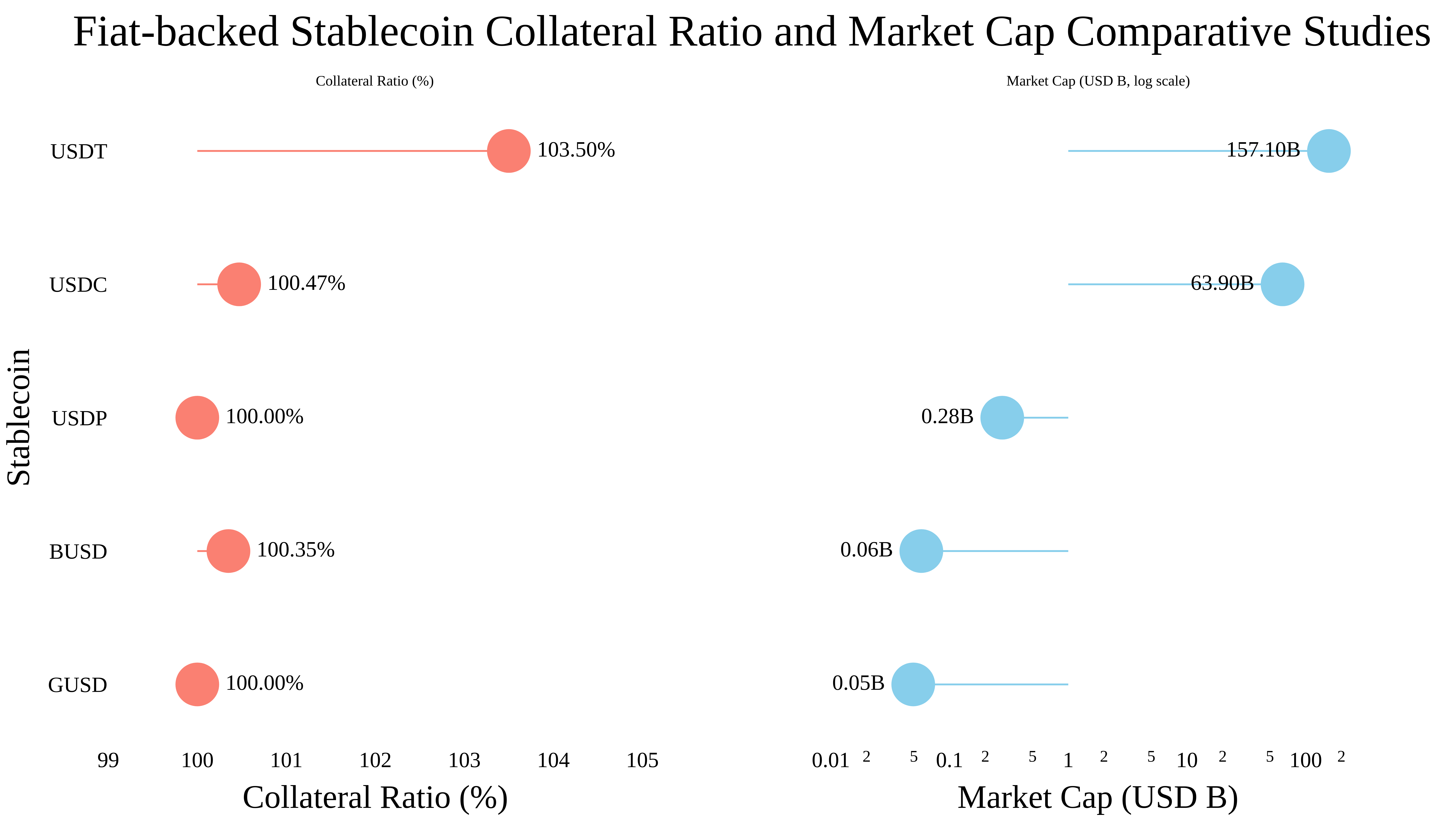}
    \caption{Collateral ratios and market capitalizations for major fiat-backed stablecoins. Market cap is plotted on a logarithmic scale. Marker size reflects prominence; lollipop stems trace each metric from baseline. Data Sources: transparency reports as indicated in Table~\ref{tab:collateral_comparison}.}
    \label{fig:stablecoin_lollipop}
\end{figure}

Figure~\ref{fig:stablecoin_lollipop} provides a comparative visualization of the reported collateral ratios and market capitalizations. Although all reported collateral ratios indicate sufficient reserves, their interpretability is contingent on the liquidity of reserve components and the credibility of the institutions providing the attestations. SEC-regulated structures, such as Circle’s reserve fund, offer standardized oversight and asset transparency. Conversely, tokens backed by a heterogeneous mix of financial instruments or attested by non-U.S. auditors may carry additional interpretive risk. The effectiveness of the collateral ratio as a signal of financial solvency therefore depends not only on the numerical sufficiency of reserves but also on the quality, auditability, and jurisdictional oversight of those reserves.

Crypto-backed stablecoins such as DAI differ fundamentally from fiat-backed models in their reliance on decentralized and overcollateralized reserve mechanisms. DAI is issued by the Maker Protocol—now governed under the Sky brand, the successor to MakerDAO—and is collateralized by on-chain assets like ETH and USDC. To maintain the peg to USD, users must deposit assets exceeding the value of minted DAI, with minimum collateralization ratios starting at 130\% depending on the asset class~\cite{makerdao_whitepaper}. However, this overcollateralization does not eliminate systemic risks. As shown by Angeris et al.~\cite{angeris2022liquidity}, decentralized liquidity markets, including those used for collateral liquidations, are susceptible to high slippage and severe liquidity constraints during price crashes. This dynamic was starkly illustrated during the March 2020 crisis, where rapid ETH price declines and network congestion led to cascading liquidations and protocol shortfalls. Thus, while DAI often exhibits a high collateral ratio, its reliance on volatile and endogenous collateral introduces fundamentally different risk exposures compared to fiat-backed stablecoins with off-chain reserves.

\subsubsection{DeFi APY}
To evaluate the interest-bearing potential of fiat-backed stablecoins, we collected yield data from decentralized finance (DeFi) protocols using the DeFiLlama yield aggregator API.\footnote{\url{https://yields.llama.fi/pools}} The data snapshot was obtained on \textit{August 3, 2025}, and includes all active on-chain liquidity pools associated with each stablecoin across multiple protocols and blockchains. Each pool record contains protocol affiliation, blockchain deployment, pool-level TVL and APY, and token metadata. A detailed explanation of the raw dataset structure is provided in Appendix Table~\ref{tab:pool_dictionary}.

The extracted metrics include: the number of DeFi protocols offering yield-bearing pools for each token, the number of blockchain networks where those pools are deployed, the minimum, maximum, and median observed annual percentage yields (APY), and the aggregated total value locked (TVL) in USD. Table~\ref{tab:stablecoin_metrics} summarizes these statistics for six leading fiat-backed stablecoins, including USDC, USDT, DAI, BUSD, USDP, and GUSD.

\begin{figure}[htbp]
    \centering
    \includegraphics[width=0.9\linewidth]{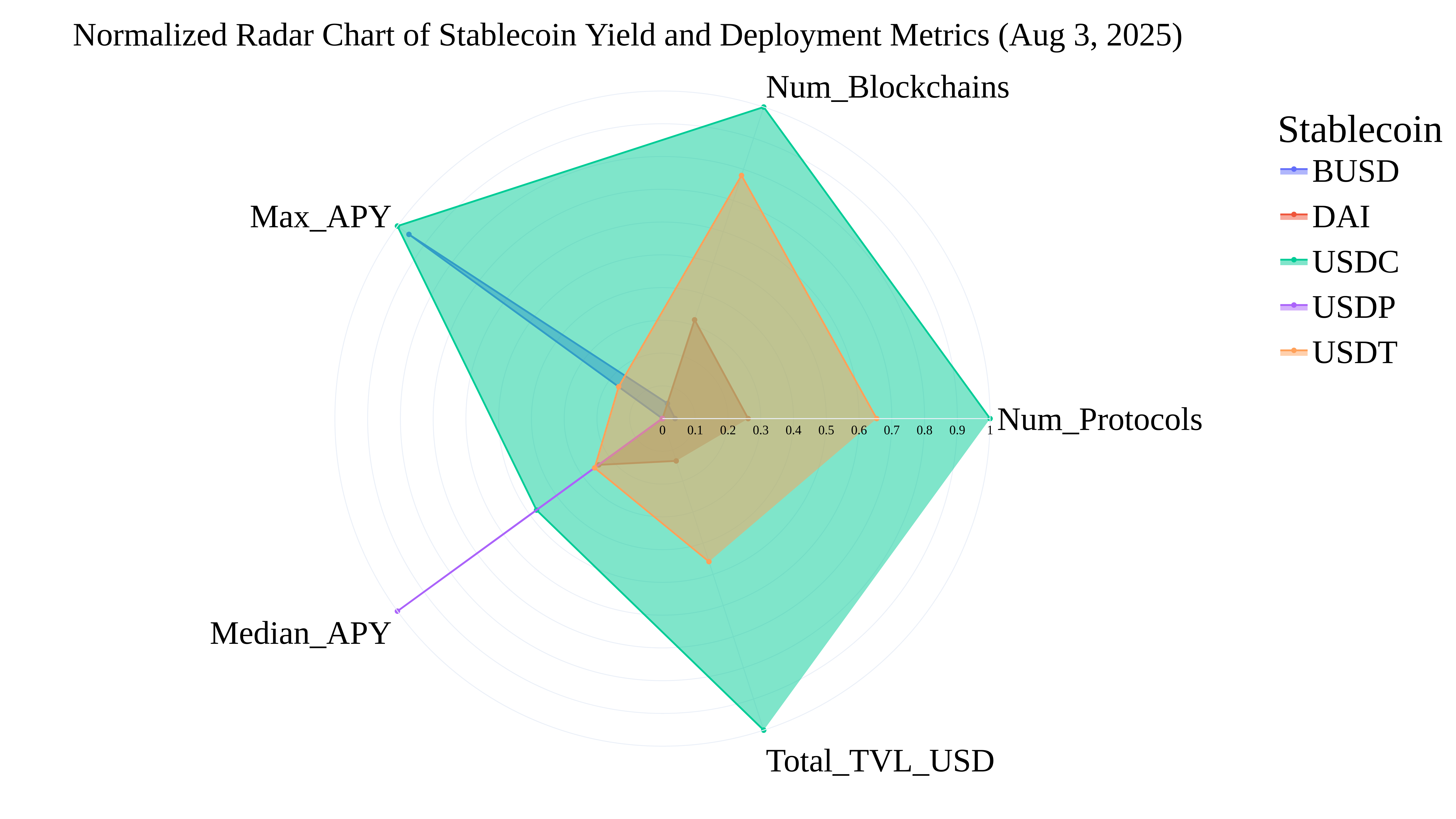}
    \caption{Radar chart comparing normalized DeFi metrics across fiat-backed stablecoins. Metrics include protocol support, blockchain deployment, maximum and median APY, and total TVL. Each metric is normalized to a 0–1 scale to enable cross-variable comparison. Data reflects the state of on-chain pools from \texttt{yields.llama.fi/pools} as of August 3, 2025.}
    \label{fig:stablecoin_radar}
\end{figure}

To facilitate comparative visualization across disparate metrics, we normalize each numerical value to the $[0, 1]$ range using min-max scaling. The resulting radar chart is presented in Figure~\ref{fig:stablecoin_radar}. This visualization allows for consistent cross-metric comparison—whereby a token scoring 1.0 in any category reflects the maximum observed value for that dimension (e.g., highest TVL, widest protocol adoption), while others are scaled proportionally.

The radar plot reveals several key insights. First, USDC exhibits the most comprehensive deployment across the DeFi landscape: it is supported by \textbf{131 protocols}, active on \textbf{42 blockchains}, and has accumulated a total TVL of over \textbf{\$5.0 billion}, alongside a strong median APY of \textbf{6.95\%}. USDT and DAI also demonstrate substantial integration, with high TVLs exceeding \textbf{\$2.2 billion} and \textbf{\$679 million}, respectively, and a presence across dozens of protocols and chains.

In contrast, tokens such as USDP and BUSD show more limited adoption. USDP is active in only a single protocol and blockchain but yields a comparatively high median APY of \textbf{14.40\%}, suggesting concentrated, possibly incentive-driven opportunities. BUSD, while more broadly deployed than USDP, reflects low effective yield and reduced DeFi traction. GUSD registers no active yield pools in the observed snapshot, indicating a complete absence or temporary dormancy in the DeFi ecosystem at the time of query.

These findings highlight the dual importance of breadth and efficiency in stablecoin yield strategies. While aggregate TVL and adoption serve as proxies for liquidity and accessibility, yield dispersion—particularly median APY—offers a measure of the stablecoin’s income-generating potential. Tokens with high APYs but narrow deployment may reflect promotional campaigns or concentrated strategies, but may lack the robustness of deeply integrated alternatives like USDC or USDT. Careful evaluation of both yield efficiency and adoption scale is therefore critical for informed DeFi participation.

\section{Applications: A Case Study in Real World Asset Tokenizations}
\label{sec:applications}

As shown in Figure~\ref{fig:eth-defi-apy-tvl}, among all Real World Asset (RWA) protocols, Maple Finance holds the highest USDC-denominated total value locked (TVL) on Ethereum. Based on a snapshot taken on August 4, 2025 at 02:29 UTC from the DeFiLlama Yield Aggregator API~\cite{defillama-api}, Maple’s largest USDC-backed DeFi pool reported a TVL of approximately \$1.94 billion. This pool delivered a composite annual percentage yield (APY) of 9.19\%, comprising a 6.99\% base APY generated from lending and credit delegation, and a 2.20\% reward APY distributed in Syrup Token (SYRUP). The associated reward contract is deployed at \texttt{0x643C...2d66}\footnote{\scriptsize Full reward token address: \texttt{0x643C4E15d7d62Ad0aBeC4a9BD4b001aA3Ef52d66}}, and the stablecoin utilized is the canonical USDC token (\texttt{0xa0b8...eb48}\footnote{\scriptsize Full USDC token address: \texttt{0xa0b86991c6218b36c1d19d4a2e9eb0ce3606eb48}}).

Maple exemplifies a hybrid DeFi–CeFi architecture in which decentralized capital is programmatically allocated to real-world borrowers through under-collateralized credit facilities~\cite{maple-docs, schar2021defi, spglobal2023credit}. Unlike traditional over-collateralized lending systems prevalent in DeFi, Maple enables under-collateralization by integrating off-chain Know Your Customer (KYC) checks, legal agreements, and credit risk assessment frameworks. Borrowers are vetted through institutional-grade underwriting processes that account for their financial identity, creditworthiness, and business fundamentals~\cite{maple-docs}. This off-chain validation permits more capital-efficient lending and bridges blockchain-native liquidity with real-world credit infrastructures.

Concurrently, lenders interact directly with Maple’s smart contracts, which automate fund allocation, interest distribution, and token issuance. The base APY is programmatically encoded within smart contracts tied to the lending pool, while additional yield—such as protocol incentives or staking rewards—is distributed via reward tokens~\cite{spglobal2023credit, maple-docs}. This composable structure exemplifies the principle of cryptographic enforcement: while trust anchors like KYC and legal documentation are centralized, capital flows and yield disbursements remain verifiable and tamper-resistant on-chain. As a result, Maple facilitates programmable credit markets with transparent and auditable risk exposure.

According to the RWA.xyz analytics dashboard~\cite{rwa-xyz}, Maple ranks as the second-largest tokenized private credit protocol, having originated over \$3.85 billion in loans as of August 4, 2025. The average base APY stood at 9.42\%, illustrating the yield opportunities made possible through smart contract automation and tokenized credit underwriting. The composability of USDC and transparency of Ethereum further enable a programmable and modular credit infrastructure, reducing friction across origination, disbursement, and repayment processes.

To illustrate Maple’s on-chain logic, we analyze two Ethereum-based case studies: one deposit and one redemption transaction. The full token flow, addresses, and quantities were parsed using Web3.py and Alchemy’s archival RPC endpoint, which enables direct retrieval of smart contract events and logs.

\begin{figure}[ht]
  \centering
  \includegraphics[width=\linewidth]{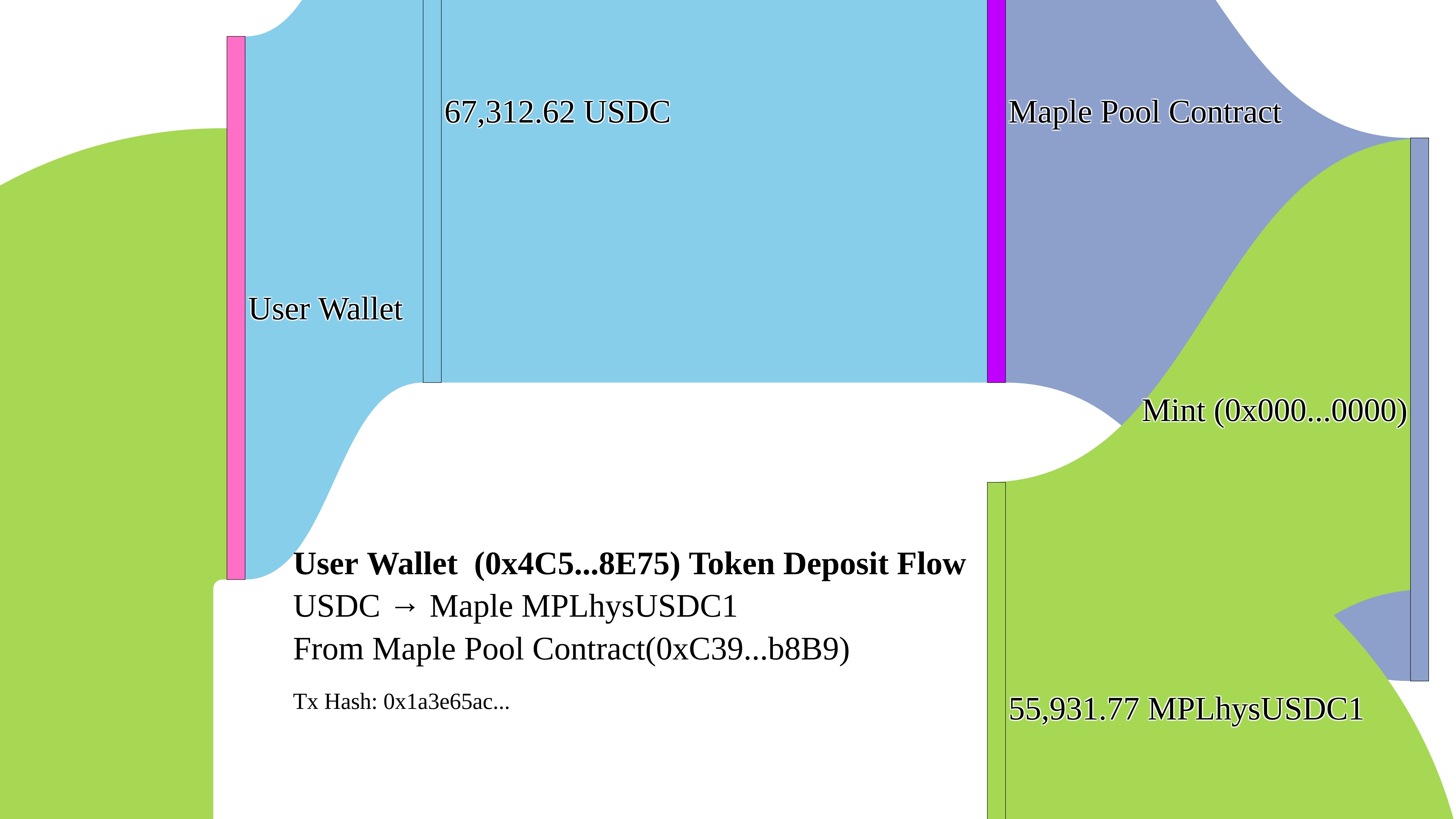}
  \caption{USDC deposit into Maple’s \texttt{MPLhysUSDC1} pool. Transaction parsed using Web3.py and Alchemy. Flows are labeled with contract roles and USD values.}
  \label{fig:maple_usdc_deposit_sankey}
\end{figure}

\paragraph{Deposit Case Study.}
In transaction \texttt{0x1a3e...498d}\footnote{\scriptsize Full Tx hash: \texttt{0x1a3e65acd3e12df20afa1985042a3bdacdff8e364ee4ebdc1b7f3e2ee3d2498d}}, the user wallet \texttt{0x4C55...8E75} deposited 67,312.62 USDC into the Maple pool \texttt{MPLhysUSDC1}, receiving 55,931.77 MPLhysUSDC1 tokens in return—representing their pro-rata share of the lending vault. The flow is visualized in Figure~\ref{fig:maple_usdc_deposit_sankey}, which includes annotated token roles, addresses, and transaction values. Off-chain, this capital is allocated to vetted institutional borrowers through Maple’s credit marketplace.

\begin{figure}[ht]
  \centering
  \includegraphics[width=\linewidth]{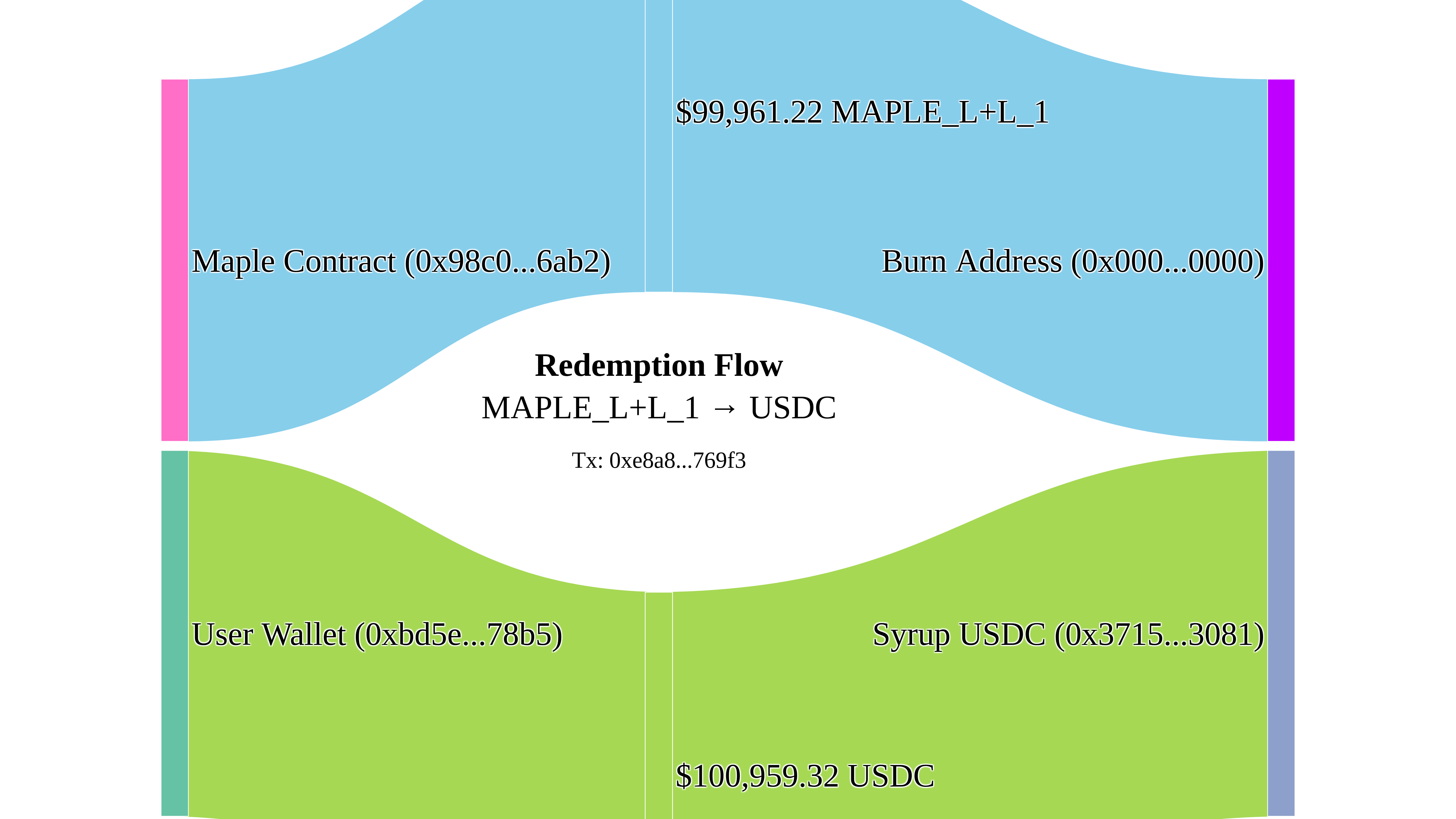}
  \caption{USDC redemption from Maple. 99,961.22 MAPLE\_L+L\_1 tokens burned, 100,959.32 USDC returned to user. Structured from on-chain logs using Web3.py.}
  \label{fig:maple_usdc_redemption_sankey}
\end{figure}

\paragraph{Redemption Case Study.}
In transaction \texttt{0xe8a8...69f3}\footnote{\scriptsize Full Tx hash: \texttt{0xe8a8772c3c3d6c2e4ff97af55ad4a2f8116245cd35361328dd65afe61cc769f3}}, a redemption was initiated where 99,961.22 \texttt{MAPLE\_L+L\_1} tokens were burned by the Maple contract at \texttt{0x98c0...6Ab2}, followed by the disbursement of 100,959.32 USDC to the user wallet \texttt{0xbd5e...78b5}. This bi-directional asset transformation is depicted in Figure~\ref{fig:maple_usdc_redemption_sankey}, representing the completion of a lending cycle backed by real-world assets and repaid on-chain.

These case studies demonstrate how stablecoins such as USDC serve as foundational instruments for enabling programmable credit flows within hybrid trust architectures. These architectures integrate off-chain components—such as compliance, legal enforcement, and credit assessment—with on-chain mechanisms for automation, auditability, and settlement finality. By functioning as fiat-referenced, interoperable digital assets, stablecoins bridge the institutional reliability of CeFi with the composability and transparency of DeFi. This synthesis facilitates the construction of capital markets that are not only scalable and globally accessible, but also designed to support financial inclusion through programmable and verifiable monetary infrastructure.

\section{Limitations and Future Research}
\label{sec:future}

This study adopts a macro-level, interdisciplinary lens to synthesize the design principles, performance metrics, and socio-technical implications of stablecoins. While this systemic view provides a broad foundation for understanding architectural trade-offs and institutional challenges, it necessarily abstracts from project-level heterogeneity and stakeholder-specific behaviors. This limitation underscores the need for future research that complements macro-level analysis with micro-level evaluations aligned to multi-objective design goals: operational efficiency, distributional inclusion, and systemic equity.

\textbf{Direction 1: Stakeholder Multi-Objectives} Table~\ref{tab:stakeholders} presents a stakeholder taxonomy segmented into four core categories: \textit{protocol developers}, \textit{supporting infrastructure}, \textit{market participants}, and \textit{regulators}. Future research should leverage this framework to empirically assess how these actors interact and negotiate competing objectives such as stability, accessibility, and legitimacy. For instance, inclusion-oriented metrics—such as global user coverage, regulatory access, or demographic disparities—require disaggregated modeling of user behavior and participation~\cite{xiao2024centralized}. Equity concerns similarly call for fine-grained tracking of validator and liquidity provider (LP) reward flows across temporal and network dimensions~\cite{yan2025data}. Recent studies on the blockchain trilemma offer structured approaches to evaluate trade-offs among decentralization, scalability, and security~\cite{fu2024quantifying,liu2022empirical,fu2023ai}. In parallel, new open-access datasets capturing daily on-chain activity~\cite{chemaya2025dataset,zhang2023blockchain,ao2022decentralized,zhang2022sok} and global sentiment trends~\cite{chen2024global} provide the empirical substrate for such analysis.

\textbf{Direction 2: Expanding Dimensions of Metrics} While this study emphasizes transparent and broadly applicable metrics, several essential dimensions remain underexplored and require project-specific modeling approaches. Governance quality, for example, is difficult to assess uniformly due to variation in institutional design. Metrics such as voting entropy, delegation concentration, and participation asymmetry must be tailored to protocol-specific contexts~\cite{liu2024economics,quan2024decoding}. Liquidity depth in decentralized markets varies based on AMM architectures and LP incentives, demanding simulation models built on high-resolution and open datasets~\cite{chemaya2025dataset}. Global accessibility, while conceptually central, lacks standardized indices and must account for jurisdictional constraints, platform access, and infrastructural disparities. Similarly, interoperability must be captured along both technical (e.g., bridge reliability, messaging protocols) and institutional (e.g., trust assumptions) lines. Addressing these gaps will require integration of structured, multi-modal metadata pipelines and adherence to emerging standards in machine learning–ready data design~\cite{akhtar2024croissant}.

\textbf{Direction 3: AI Evaluation and Multi-Modal Data} The Web3 ecosystem—characterized by transparent execution, algorithmic governance, and high-frequency behavioral data—offers a novel empirical environment for evaluating artificial intelligence, including foundation models and agentic systems. Stablecoins, in particular, serve as dynamic platforms for testing agent behaviors, incentive responses, and coordination strategies. These systems enable new paradigms for AI evaluation focused on interpretability, decentralized interaction, and learning in strategic settings~\cite{zhang2023machine,fu2024dam,chen2024finml,zhang2023understand}. Reinforcement learning experiments in decentralized liquidity provision and DAO governance offer real-world opportunities for mechanism design~\cite{tian2024redesign}. To fully realize this potential, future research must prioritize the development of multi-modal, high-quality, and openly accessible benchmarks. Integrating on-chain transaction data with off-chain contextual information—such as demographic, legal, and institutional metadata—will support broader goals in global economic modeling and inclusive digital finance~\cite{zhang2023future,wu2024trust,chen2024global}.

In summary, this study establishes a foundational framework for analyzing stablecoin systems at scale, while identifying key limitations that open promising directions for future research. Bridging macro-level synthesis with micro-level, stakeholder-informed inquiry—alongside expanding metric frameworks and leveraging the stablecoin ecosystem for AI benchmarking—will be essential for building resilient, inclusive, and scientifically robust programmable monetary systems.

\begin{acks}

This research has benefited from insightful conversations at two academic venues: the 2025 ADEF-Xueshuo Summer Institute on \textit{Financial Economics Meets Data Science}, held July 7–9, 2025 at Xiamen University, and the China Computer Federation (CCF) Young Computer Scientist and Engineer Forum (YOCSEF) Event on \textit{Stablecoins and the Future of International Financial Centers}, held at the University of Hong Kong on July 12, 2025. We thank participants from both venues for their thoughtful engagement and constructive feedback.
\end{acks}

\bibliographystyle{ACM-Reference-Format}
\input{main.bbl}

\section{Additional Tables and Figures}
\input{tables/taxonomy}
\input{tables/data.tex}
\input{tables/pipeline}
\input{tables/stability}
\input{tables/collateral}

\input{tables/yield}

\input{tables/stakeholders}
\appendix
\setcounter{section}{0}
\input{tables/policy}
\input{tables/llama}
Table~\ref{tab:pool_dictionary} provides a comprehensive dictionary of variables used in the \texttt{yields.llama.fi/pools} dataset. As an example, consider the following single pool observation from Lido’s stETH staking pool on Ethereum obtained on August 3, 2025. At the time of query, the \texttt{apy} was reported as 2.632\%, identical to the \texttt{apyBase} since no external rewards (\texttt{apyReward}) were present. The pool had an impressive \texttt{tvlUsd} of over $30.8$ billion, suggesting broad adoption and trust. According to the ML-based forecast, the \texttt{predictedClass} was ``Stable/Up'' with a 74\% confidence score (\texttt{predictedProbability}). This suggests that Lido’s staking yield is statistically stable, consistent with its low volatility (\texttt{sigma = 0.052}) and large historical sample size (\texttt{count = 1157}). This single observation illustrates how such data can support protocol comparisons and yield reliability assessments.

\end{document}

%% file: main.bbl

%% file: tables/taxonomy.tex
\definecolor{facetgray}{gray}{0.95}
\onecolumn
\begin{table}[htbp!]
\caption{\textbf{Stablecoin Design Taxonomy: Design Facets, Features, and Examples}}
\label{tab:taxonomy}
\renewcommand{\arraystretch}{1.3}
\setlength{\tabcolsep}{4pt}
\centering
\begin{tabular}{@{}l p{5.5cm} l l@{}}
\toprule
\textbf{Design Facet} & \textbf{Feature} & \textbf{Design Option} & \textbf{Example} \\
\midrule

\multirow{5}{*}{Stabilization Mechanism} 
& \multirow{5}{=}{\begin{tabular}[t]{@{}l@{}}Price Stability \\ \small $\mathbb{E}\left[ d(P_t, P_t^{\text{peg}}) \right] < \varepsilon$\end{tabular}} 
& Fiat-backed & USDC~\cite{circleUSDC} \\
& & Crypto-backed & DAI~\cite{makerdaoWhitepaper} \\
& & Algorithmic(+Crypto-backed) & FRAX~\cite{fraxDocs} \\
& & Synthetic & USDe~\cite{ethenaDocs} \\
& & Commodity-backed & PAXG~\cite{paxgWhitepaper} \\
\cmidrule(l){1-4}

\multirow{4}{*}{Custodial Structure}
& \multirow{4}{=}{\begin{tabular}[t]{@{}l@{}}Distributed Trust \\ \small $\left| \frac{\partial P_t}{\partial T_i} \right| \ll 1 \quad \forall i$\end{tabular}}
& Centralized & USDC~\cite{circleUSDC} \\
& & On-chain collateral & DAI~\cite{makerdaoWhitepaper} \\
& & Algorithmic reserves & USDe~\cite{ethenaDocs} \\
& & Hybrid & FRAX~\cite{fraxDocs} \\
\cmidrule(l){1-4}

\multirow{4}{*}{Interest Mechanism}
& \multirow{4}{=}{\begin{tabular}[t]{@{}l@{}}Interest Generation \\ \small $\mathbb{E}[r_t^{\text{stable}}] \geq r_t^{\text{risk-free}} + \eta$ \end{tabular}}
& No yield & USDC (classic)~\cite{circleUSDC} \\
& & RWA-based yield & USDC (enhanced yield)~\cite{circleUSDC}\\
& & Crypto-collateralized yield & DAI (via sDAI)~\cite{makerdaoWhitepaper} \\
& & Hybrid/synthetic yield & USDe (via sUSDe)~\cite{ethenaDocs} \\
\cmidrule(l){1-4}

\multirow{2}{*}{Market Access}
& \multirow{2}{=}{\begin{tabular}[t]{@{}l@{}}Global Accessibility \\ \small $\Pr_{u \sim \mathcal{U}}[\exists p \in \{\text{CEX}, \text{DEX}\} : A(u,t,p) = 1] \rightarrow 1$\end{tabular}}
& Single-region CEX listing & PYUSD~\cite{paypal2023pyusd} \\
& & Multi-CEX global listing & USDT~\cite{tetherWhitepaper} \\
\cmidrule(l){1-4}

\multirow{3}{*}{Governance Model}
& \multirow{3}{=}{\begin{tabular}[t]{@{}l@{}}Distributed Trust, Programmability \\ \small $f: \mathcal{E} \rightarrow \mathcal{A},\ f(e)\text{ executes on-chain}$\end{tabular}}
& Corporate & USDC~\cite{circleUSDC} \\
& & DAO & DAI~\cite{makerdaoWhitepaper} \\
& & Hybrid & FRAX~\cite{fraxDocs} \\
\cmidrule(l){1-4}

\multirow{2}{*}{Protocol Interoperability}
& \multirow{2}{=}{\begin{tabular}[t]{@{}l@{}}Composability \\ \small $\forall \pi_j \in \mathcal{P},\ C(P_t, \pi_j) = 1$\end{tabular}}
& Single-chain integration & PYUSD~\cite{paypal2023pyusd} \\
& & Multi-chain & USDC~\cite{circleUSDC} \\
\cmidrule(l){1-4}

\multirow{2}{*}{Liquidity Mechanism}
& \multirow{2}{=}{\begin{tabular}[t]{@{}l@{}}Liquidity (Slippage Control) \\ \small $\delta_t = \frac{|P_t^{\text{ask}} - P_t^{\text{bid}}|}{P_t^{\text{peg}}} \ll 1$\end{tabular}}
& CEX double auction & USDT~\cite{tetherWhitepaper} \\
& & AMM (DEX pool) & DAI~\cite{makerdaoWhitepaper} \\\\
\bottomrule
\end{tabular}

\vspace{0.5em}
\begin{minipage}{\linewidth}
\footnotesize
\justifying

\textbf{Formula Notation:} \\
$P_t$: stablecoin market price at time $t$;  
$P_t^{\text{peg}}$: peg price (e.g., 1 USD);  
$\delta_t$: relative bid-ask spread;  
$r_t^{\text{stable}}$: yield rate of the stablecoin;  
$r_t^{\text{risk-free}}$: benchmark fiat rate;  
$T_i$: trust component (e.g., oracle, multisig);  
$A(u,t,p)$: access of user $u$ at time $t$ on platform $p$;  
$\mathcal{E}$: on-chain events;  
$\mathcal{A}$: action space;  
$\mathcal{P}$: set of DeFi protocols;  
$C(P_t, \pi_j)$: composability indicator with protocol $\pi_j$. \\

\textbf{Facet Explanation (with Definitions)}: \\

\emph{Stabilization Mechanism} refers to the protocol's approach for maintaining the stablecoin's target peg (usually $P_t^{\text{peg}} = \$1$). This can involve fiat reserves (e.g., USDC~\cite{circleUSDC}), crypto-collateralized smart contracts (DAI~\cite{makerdaoWhitepaper}), algorithmic supply adjustments (FRAX~\cite{fraxDocs}), or synthetic hedging (USDe~\cite{ethenaDocs}). The effectiveness is often measured via mean absolute deviation (MAD). Commodity-backed assets like PAXG~\cite{paxgWhitepaper}, XAUT~\cite{xautWhitepaper}, and KAG~\cite{kinesisDocs} extend the definition of “stablecoin” to include real-world physical anchors, enabling blockchain-based exposure to precious metals. \\

\emph{Custodial Structure} captures the control and transparency of the reserve assets. Custody types include centralized (USDC), on-chain smart contracts (DAI), algorithmic automation (USDe), or hybrids combining governance and contracts (FRAX). \\

\emph{Interest Mechanism} defines whether and how yield is passed to users. Post-2021 monetary tightening enabled yield-bearing stablecoins via exposure to real-world assets (RWAs) like U.S. Treasury bills~\cite{usdnarrative}. Issuers either distribute yield or retain a net-interest margin. Crypto-backed (sDAI~\cite{makerdaoWhitepaper}) and synthetic forms (sUSDe~\cite{ethenaDocs}) offer alternatives to RWA-based models. \\

\emph{Market Access} denotes how globally users can acquire and trade the stablecoin. Access channels include centralized exchanges (CEXs), decentralized exchanges (DEXs), or both. \\

\emph{Governance Model} distinguishes who holds control over upgrades and operations. This ranges from corporate entities (e.g., Circle), decentralized autonomous organizations (DAOs), or hybrid models (e.g., FRAX AMOs~\cite{fraxDocs}). Governance is often programmable via smart contracts mapping events to actions, $f: \mathcal{E} \rightarrow \mathcal{A}$. \\

\emph{Protocol Interoperability} measures the degree of composability with other protocols. Multi-chain operability (e.g., USDC~\cite{circleUSDC}) enhances utility, whereas single-chain systems like PYUSD~\cite{paypal2023pyusd} are more constrained. \\

\emph{Liquidity Mechanism} represents how users buy/sell the stablecoin and how price discovery occurs. Centralized limit order books (CEX double auction) or on-chain AMMs (Automated Market Makers) govern slippage $\delta_t$ and execution certainty. \\

\textbf{Abbreviations}: \\
\textbf{CEX}: Centralized Exchange; 
\textbf{DEX}: Decentralized Exchange; 
\textbf{RWA}: Real-World Asset; 
\textbf{AMM}: Automated Market Maker; 
\textbf{DAO}: Decentralized Autonomous Organization; 
\textbf{TMMF}: Tokenized Money Market Fund

\end{minipage}
\end{table}

%% file: tables/data.tex
\begin{table}[htbp!]
\caption{Stablecoin Metrics Data Types, Sources, and Query Interfaces}
\label{tab:stablecoin_data_sources}
\centering
\small
\renewcommand{\arraystretch}{1.6}
\begin{tabular}{@{}p{4.2cm} p{2.2cm} p{4.2cm} p{4.2cm}@{}}
\toprule
\textbf{Design Feature / What It Measures} 
& \textbf{Data Type} 
& \textbf{\textcolor{firsthand}{\faTools}~First-Hand Source (Original)} 
& \textbf{\textcolor{secondhand}{\faSatelliteDish}~Second-Hand Source (API)} \\
\midrule

Price stability and volatility 
& \textcolor{offchain}{\faCircle\ Off-chain} 
& Binance/Uniswap trades~\cite{binanceAPI, uniswapSubgraph} 
& CoinGecko~\cite{coingeckoAPI}, CoinMetrics~\cite{coinmetrics2024}, Bitquery~\cite{bitqueryDocs} \\

Collateralization (trust)
& \textcolor{hybrid}{\faAdjust\ Hybrid}
& Circle reports~\cite{circleUSDC}, MakerDAO contracts~\cite{makerdaoWhitepaper}
& Chainlink PoR~\cite{chainlinkDocs}, Bitquery~\cite{bitqueryDocs} \\

DeFi yield and APY 
& \textcolor{onchain}{\faCube\ On-chain} 
& Aave/Compound via ethers.js~\cite{ethersDocs}, Web3.js~\cite{web3jsDocs} 
& The Graph~\cite{thegraphDocs}, Dune~\cite{duneAnalytics} \\

Exchange reach and user base 
& \textcolor{offchain}{\faCircle\ Off-chain} 
& Binance API~\cite{binanceAPI}, SEC~\cite{secEDGAR} 
& CoinGecko exchange stats~\cite{coingeckoAPI} \\

Governance distribution 
& \textcolor{onchain}{\faCube\ On-chain} 
& Smart contract events via Infura~\cite{infuraDocs}, Alchemy~\cite{alchemyDocs} 
& Snapshot~\cite{snapshotAPI}, Tally~\cite{tallyAPI} \\

Interoperability (bridging) 
& \textcolor{hybrid}{\faAdjust\ Hybrid}
& LayerZero~\cite{layerzeroDocs}, Wormhole~\cite{wormholeDocs} bridge explorer data 
& Chainlist multi-chain registry~\cite{chainlist2024} \\

Liquidity depth (future) 
& \textcolor{onchain}{\faCube\ On-chain}
& Uniswap/Curve LP state~\cite{curveSubgraph}
& 0x API~\cite{0xAPI} \\
\bottomrule
\end{tabular}
\end{table}

%% file: tables/pipeline.tex
\begin{table}[htbp]
\caption{Metric Computation Workflow}
\label{tab:metric_pipeline}
\centering
\small
\renewcommand{\arraystretch}{1.4}
\begin{tabular}{@{}p{1.2cm} p{3.4cm} p{9.0cm}@{}}
\toprule
\textbf{Step} & \textbf{Stage} & \textbf{Operation Description} \\
\midrule
\textcolor{firsthand}{\faDownload} & \textbf{Extract} & Collect raw streams (e.g., prices \(P_t\), reserves \(C_T\), supplies \(S_T\), rates \(r_{i,T}\)) via APIs or blockchain tools (ethers.js, Infura). \\
\textcolor{hybrid}{\faSlidersH} & \textbf{Normalize} & Align timestamps, resample to fixed resolution \(\tau\), and handle missing entries. \\
\textcolor{onchain}{\faCalculator} & \textbf{Compute} & Apply Equations~\ref{eq:rmse}–\ref{eq:access} to derive metric scores. \\
\textcolor{secondhand}{\faSave} & \textbf{Export} & Output normalized data and visualization charts. \\
\bottomrule
\end{tabular}
\end{table}

%% file: tables/stability.tex
\definecolor{tightpeg}{RGB}{198,239,206}      
\definecolor{moderatepeg}{RGB}{255,235,156}   
\definecolor{widepeg}{RGB}{255,199,206}       

\definecolor{tightpegtext}{RGB}{0,100,0}      
\definecolor{moderatepegtext}{RGB}{102,85,0}  
\definecolor{widepegtext}{RGB}{153,0,0}       

\begin{table}[htbp]
\centering
\caption{Observed Deviations from Peg for Stablecoins and Tokenized Assets on 2025-07-15}
\label{tab:peg_deviations}
\scriptsize
\renewcommand{\arraystretch}{1.2}
\begin{tabular}{l l l r r r}
\toprule
\textbf{Asset} & \textbf{Design Type} & \textbf{Peg Reference} & \textbf{Peg Value (USD)} & \textbf{Observed Price (USD)} & \textbf{Deviation (\%)} \\
\midrule
\rowcolor{tightpeg} USDC  & Fiat-backed Stablecoin   & USD  & 1.0000    & 0.999786     & -0.0214 \\
\rowcolor{tightpeg} DAI   & Crypto-backed Stablecoin & USD  & 1.0000    & 1.000271     & +0.0271 \\
\rowcolor{tightpeg} USDT  & Fiat-backed Stablecoin   & USD  & 1.0000    & 1.000178     & +0.0178 \\
\rowcolor{widepeg}  PAXG  & Gold-backed Token        & Gold & 3297.30   & 3338.037825  & +1.2355 \\
\rowcolor{widepeg}  XAUT  & Gold-backed Token        & Gold & 3297.30   & 3324.242832  & +0.8171 \\
\rowcolor{moderatepeg} BUSD  & Fiat-backed Stablecoin   & USD  & 1.0000    & 0.998839     & -0.1161 \\
\rowcolor{tightpeg} GUSD  & Fiat-backed Stablecoin   & USD  & 1.0000    & 1.000042     & +0.0042 \\
\rowcolor{moderatepeg} PAX   & Fiat-backed Stablecoin   & USD  & 1.0000    & 0.997704     & -0.2296 \\
\bottomrule
\end{tabular}

\vspace{0.8em}
\begin{minipage}{0.94\linewidth}
\footnotesize
\textbf{Table Notes:}

\textbf{Peg Definition and Source:} Peg values (\textit{Peg Value (USD)}) reflect the intended reference price. Fiat stablecoins are pegged to \(\$1.00\) USD. Gold-backed tokens use a derived spot price of gold on 2025-07-15 from:
\[
\text{USD/oz} = (\text{PLN/g}) \times (\text{USD/PLN}) \times 31.1035
\]
via the NBP APIs: \url{http://api.nbp.pl/}.

\textbf{Observed Price Data:} Retrieved from CoinMetrics Community API (\url{https://coinmetrics.io/community-api/}) using the \texttt{PriceUSD} (VWAP) metric:
\[
\text{VWAP} = \frac{\sum_{i=1}^{N} P_i \cdot V_i}{\sum_{i=1}^{N} V_i}
\]

\textbf{Deviation Metric:}
\[
\text{Deviation (\%)} = \left( \frac{\text{Observed Price} - \text{Peg Value}}{\text{Peg Value}} \right) \times 100
\]

\textbf{Row Color Legend (based on absolute deviation):}
\begin{itemize}[leftmargin=1.6em]
  \item \textcolor{tightpegtext}{\textbf{Tight Peg (\(\leq 0.05\%\))}:} Peg well maintained. Reflects highly liquid and mature backing.
  \item \textcolor{moderatepegtext}{\textbf{Moderate Deviation (0.05--0.5\%)}:} Slight deviation from peg, possibly due to low liquidity or minor volatility.
  \item \textcolor{widepegtext}{\textbf{Significant Deviation (\(> 0.5\%\))}:} Market-based instruments (e.g., gold tokens) show natural volatility.
\end{itemize}
\end{minipage}
\end{table}

%% file: tables/collateral.tex
\definecolor{cr_green}{RGB}{0,128,0}
\definecolor{cr_blue}{RGB}{0,0,180}
\definecolor{cr_red}{RGB}{180,0,0}
\definecolor{tres}{RGB}{240,248,255} 
\definecolor{cash}{RGB}{255,250,205} 
\definecolor{mixed}{RGB}{255,228,225} 

\begin{table}[htbp]
\caption{Summary of Fiat-Backed Stablecoin Collateralization (Latest Reports)}
\label{tab:collateral_comparison}
\centering
\scriptsize
\renewcommand{\arraystretch}{1.2}
\setlength{\tabcolsep}{3pt}
\begin{tabular}{@{}>{\bfseries\color{blue}}l>{\color{black}}l>{\color{black}}r>{\color{black}}r>{\color{black}}l>{\color{black}}>{\centering\arraybackslash}p{2.3cm}>{\color{black}}p{2.6cm}>{\color{black}}l>{\color{black}}l@{}}
\toprule
Asset & Date & Market Cap & Reserves & \textbf{CR (\%)} & Reserve Type & Regulator / Auditor & Issuer & Ref. \\
\midrule
USDC & Jul 31, 2025 & \$63.9B & \$64.2B & \textcolor{cr_blue}{100.47} & \cellcolor{tres}U.S. Treasuries, Overnight Repo & SEC / BlackRock & Circle & \cite{circle2025} \\
USDT & Jun 30, 2025 & \$157.1B & \$162.6B & \textcolor{cr_green}{103.50} & \cellcolor{mixed}T-Bills, Cash, Gold, Bitcoin & BDO Italia & Tether & \cite{tether2025} \\
GUSD & May 30, 2025 & \$49.4M & \$49.4M & \textcolor{cr_blue}{100.00} & \cellcolor{cash}Cash Equivalents & BPM LLP & Gemini & \cite{gusd2025} \\
BUSD & Jun 30, 2025 & \$57.8M & \$58.0M & \textcolor{cr_blue}{100.35} & \cellcolor{cash}Cash Equivalents & Withum & Paxos & \cite{busd2025} \\
FAX (USDP) & Jun 30, 2025 & \$277.8M & \$277.8M & \textcolor{cr_blue}{100.00} & \cellcolor{cash}Cash Equivalents & Withum & Paxos & \cite{usdp2025} \\
\bottomrule
\end{tabular}

\vspace{0.8em}
\begin{minipage}{0.9\linewidth}
\footnotesize
\textbf{Notes:} The Collateral Ratio (CR) is computed as the ratio of reserves to market cap. 
\textcolor{cr_green}{Green} indicates overcollateralization ($>$100\%), 
\textcolor{cr_blue}{blue} is fully collateralized (100\%), and 
\textcolor{cr_red}{red} would indicate undercollateralization (not present here).

Reserve Type highlights: 
\cellcolor{tres}blue = Treasuries, 
\cellcolor{cash}yellow = cash equivalents, 
\cellcolor{mixed}pink = mixed assets like crypto, gold, and cash. 
These categories inform the transparency and risk profiles of stablecoins.
\end{minipage}
\end{table}

%% file: tables/yield.tex
\definecolor{highbg}{RGB}{230, 255, 230}     
\definecolor{mediumbg}{RGB}{255, 255, 230}   
\definecolor{lowbg}{RGB}{255, 230, 230}      

\definecolor{hightext}{RGB}{0, 120, 0}       
\definecolor{mediumtext}{RGB}{180, 130, 0}   
\definecolor{lowtext}{RGB}{160, 0, 0}        
\definecolor{graytext}{gray}{0.3}            

\begin{table}[htbp]
\caption{Summary of Performance Metrics for Selected Fiat-Backed Stablecoins (DeFi Aggregator Snapshot)}
\label{tab:stablecoin_metrics}
\centering
\scriptsize
\renewcommand{\arraystretch}{1.2}
\begin{tabular}{@{}lrrrrrr@{}}
\toprule
\textbf{Token} & \textbf{\#Protocols} & \textbf{\#Blockchains} & \textbf{Min APY (\%)} & \textbf{Max APY (\%)} & \textbf{Median APY (\%)} & \textbf{Total TVL (USD)} \\
\midrule
\rowcolor{mediumbg} BUSD & 6   & 3  & 0.00  & 79.54 & 0.20  & \$2,036,536 \\
\rowcolor{mediumbg}   DAI  & 35  & 14 & 0.00  & 14.33 & 3.61  & \$679,747,200 \\
\rowcolor{highbg}   USDC & 131 & 42 & 0.00  & 82.50 & 6.95  & \$5,004,164,000 \\
\rowcolor{lowbg}    USDP & 1   & 1  & 14.40 & 14.40 & 14.40 & \$21,919 \\
\rowcolor{highbg}   USDT & 86  & 33 & 0.00  & 25.60 & 3.84  & \$2,297,511,000 \\
\rowcolor{lowbg}    GUSD & 0   & 0  & 0.00  & 0.00  & 0.00  & \$0 \\
\bottomrule
\end{tabular}

\vspace{0.8em}
\begin{minipage}{0.95\linewidth}
\footnotesize
\textbf{Notes:} This table presents yield-related metrics for fiat-backed stablecoins based on data aggregated from DeFiLlama (\url{https://yields.llama.fi/pools}) retrieved on \textit{August 3, 2025 at 07:19 UTC}.

\begin{itemize}
    \item \textbf{Token}: Stablecoin ticker symbol.
    \item \textbf{\#Protocols}: Count of distinct DeFi platforms offering interest-bearing pools for the stablecoin.
    \item \textbf{\#Blockchains}: Number of blockchain networks where yield-bearing pools for the token are deployed.
    \item \textbf{Min / Max / Median APY}: Observed minimum, maximum, and median annual percentage yields.
    \item \textbf{Total TVL (USD)}: Aggregated total value locked in USDC equivalent across all observed pools.
\end{itemize}

\textcolor{graytext}{\textbf{TVL Activity Tier (Color Legend):}} 
\textcolor{hightext}{\cellcolor{highbg}High ($\geq\$1B$)}, 
\textcolor{mediumtext}{\cellcolor{mediumbg}Moderate (\$1M–\$1B)}, 
\textcolor{lowtext}{\cellcolor{lowbg}Low ($<\$1M$)}.
\end{minipage}
\end{table}

%% file: tables/stakeholders.tex
\definecolor{coreblue}{RGB}{52, 101, 164}
\definecolor{infragray}{gray}{0.3}
\definecolor{usergreen}{RGB}{60, 179, 113}
\definecolor{reggray}{RGB}{120, 120, 120}

\onecolumn

\begin{table}[htbp!]
\caption{\textbf{Stakeholder Taxonomy in the Stablecoin Ecosystem}}
\label{tab:stakeholders}
\renewcommand{\arraystretch}{1.5}
\setlength{\tabcolsep}{8pt}
\centering
\begin{tabular}{@{}p{4.5cm} p{4.8cm} p{5.7cm}@{}}
\toprule
\textbf{Category} & \textbf{Subgroup} & \textbf{Representative Example} \\
\midrule
\textbf{\textcolor{coreblue}{\faCubes}\hspace{0.5em} Core Protocol Stakeholders} 
& Issuers & Circle (\texttt{USDC})~\cite{circleUSDC} \\
& Governance Entities & MakerDAO (\texttt{DAI})~\cite{makerdaoWhitepaper} \\
& Custodians & BNY Mellon~\cite{bnymellon2021custody} \\

\textbf{\textcolor{infragray}{\faCogs}\hspace{0.5em} Supporting Infrastructure} 
& Auditors \& Attestors & Grant Thornton~\cite{grantthornton2023attest} \\
& Oracles & Chainlink~\cite{chainlinkDocs} \\
& Blockchain Platforms & Ethereum~\cite{ethereumDocs} \\
& Bridges & LayerZero~\cite{layerzeroDocs} \\

\textbf{\textcolor{usergreen}{\faChartBar}\hspace{0.5em} End-User \& Market Stakeholders} 
& Institutional Users & Franklin Templeton (tokenized fund using USDC)~\cite{franklin2023usdc} \\
& Retail Users & Global remittance users~\cite{visa2023remittance} \\
& Exchanges \& Market Makers & Binance~\cite{tetherWhitepaper} \\
& Developers \& Integrators & Aave~\cite{aaveDocs} \\
& Analytics \& Ratings & CoinMetrics~\cite{coinmetrics2025} \\

\textbf{\textcolor{reggray}{\faBalanceScale}\hspace{0.5em} Public Oversight \& Regulation} 
& Financial Regulators & NYDFS (regulates USDC and GUSD)~\cite{nydfs2022usdc} \\
& Global Standards Bodies & FSB (stablecoin recommendations)~\cite{fsb2023guidance} \\
\bottomrule
\end{tabular}

\vspace{0.8em}
\begin{minipage}{0.95\linewidth}
\footnotesize
\textbf{Table Notes:} \\
Stakeholders are categorized into four functional groups: \\
(1) \textit{\textcolor{coreblue}{Core Protocol Stakeholders}} \textcolor{coreblue}{(\faCubes)} manage issuance, governance, and custodianship of stablecoins. \\
(2) \textit{\textcolor{infragray}{Supporting Infrastructure}} \textcolor{infragray}{(\faCogs)} offers technical and middleware foundations such as oracle feeds, blockchain platforms, audits, and bridges. \\
(3) \textit{\textcolor{usergreen}{End-User and Market Stakeholders}} \textcolor{usergreen}{(\faChartBar)} include entities that hold, transact, or integrate stablecoins—including institutional users (e.g., Franklin Templeton), retail remittance users, exchanges, DeFi integrators, and analytics providers. \\
(4) \textit{\textcolor{reggray}{Public Oversight and Regulation}} \textcolor{reggray}{(\faBalanceScale)} includes public regulators (e.g., NYDFS) and global standard-setting bodies (e.g., FSB). \\
All representative examples are supported by primary sources.

\vspace{0.5em}
\textbf{Abbreviation Legend:} \\
BNY Mellon = Bank of New York Mellon; 
FSB = Financial Stability Board; 
NYDFS = New York Department of Financial Services; 
USDC = USD Coin; 
DAI = MakerDAO-issued stablecoin; 
GUSD = Gemini Dollar; 
DeFi = Decentralized Finance.

\end{minipage}
\end{table}

%% file: tables/policy.tex
clearpage
\begin{landscape}
\begin{table}
  \caption{Comparative Overview of Stablecoin Policies Across Key Jurisdictions (Sorted by Effective Date)}
  \label{tab:landscape-stablecoin}
  \centering
  \scriptsize
  \renewcommand{\arraystretch}{1.2}
  \rowcolors{2}{gray!10}{white}
  \begin{tabular}{@{}l p{2.8cm} p{3.6cm} p{4.2cm} p{3.5cm} p{2.6cm} p{1.6cm}@{}}
    \toprule
    \textbf{Jurisdiction} & \textbf{Regulatory Stance} & \textbf{Core Law / Policy} & \textbf{Issuance Rules \& Restrictions} & \textbf{Pilot Projects / Market Activity} & \textbf{Supervisory Authority} & \textbf{Effective Date} \\
    \midrule
    Switzerland & \textcolor{green!60!black}{\checkmark\ Supportive} & FINMA Guidance \cite{swiss_dlt_regulation} & Segregated reserves, AML compliance & Crypto hub: Zug, Zurich & FINMA & 2019 \\
    Japan & \textcolor{orange!80!black}{\textpm\ Flexible} & Funds Settlement Act \cite{japan_fsact2022} & Only banks/trusts with PSP license & NTT and major banks onboard & FSA & 2022 \\
    Singapore & \textcolor{orange!80!black}{\textpm\ Sandbox-Based} & PSA Amendment \cite{sg_psa2023} & Payment coins: license + 1:1; non-payment: AML only & DBS SGD token pilot & MAS & 2023 \\
    Australia & \textcolor{orange!80!black}{\textpm\ Cautious} & ASIC Reg. Guide 274 \cite{aus_fsl2023} & Licensed, fiat-backed only; algorithmic banned & CBA cross-border stablecoin & ASIC & 2023 \\
    EU & \textcolor{orange!80!black}{\textpm\ Restrictive} & MiCA Regulation \cite{eu_mica2023} & €100M/day cap, no interest, EU registration & Exchanges push USDT delisting & EC, ECB & 2024 \\
    USA & \textcolor{orange!80!black}{\textpm\ USD-Centric} & GENIUS Act (2025) \cite{us_genius_s1582_2025} & USD-peg, federal license, no algorithmic coins, daily reserve audit & USDC IPO; USDT still dominant & Fed, OCC, SEC & 2025-06 \\
    H.K. SAR & \textcolor{orange!80!black}{\textpm\ Experimental} & Stablecoin Ordinance \cite{hk_stablecoin_ordinance} & HKD 25M capital, 1:1 fiat reserve, license required & Circle and JD Tech pilots & HKMA & 2025-08 \\
    Korea & \textcolor{orange!80!black}{\textpm\ Controlled} & Digital Asset Basic Law \cite{korea_stablecoin2025} & $\ge$ KRW 500M capital, AML protection & Corporate adoption at 58\% & FSC & 2025-H2 \\
    UK & \textcolor{orange!80!black}{\textpm\ In Consultation} & Treasury Consultation \cite{uk_policy2026} & GBP-focused, AML, licensing required & BoE exploring CBDC coexistence & FCA, BoE & 2026 (planned) \\
    Mainland China & \textcolor{red!80!black}{\texttimes\ Prohibited} & PBoC directive \cite{china_pbc2025} & Private issuance banned; RMB is sole digital currency & Digital RMB rollout & PBoC, CSRC & --\footnotemark[1] \\
    Russia & \textcolor{red!80!black}{\texttimes\ Banned} & CBR Directive \cite{russia_cbdc_only} & Private stablecoins banned & Digital ruble cross-border tests & CBR & --\footnotemark[2] \\
    \bottomrule
  \end{tabular}

  \vspace{0.5em}
  {\footnotesize
  \begin{minipage}{\linewidth}
  \justifying

  \textbf{Legend:}
  \textcolor{green!60!black}{\checkmark\ Supportive} = least restricted;  
  \textcolor{orange!80!black}{\textpm\ Moderate} = sandboxed, cautious, or evolving regulation;  
  \textcolor{red!80!black}{\texttimes\ Restrictive} = prohibitionist or CBDC-exclusive.\\[0.8ex]

  \textbf{Abbreviations:}
  AML = Anti-Money Laundering,  
  BoE = Bank of England,  
  CBA = Commonwealth Bank of Australia,  
  CBR = Central Bank of Russia,  
  CSRC = China Securities Regulatory Commission,  
  EC = European Commission,  
  ECB = European Central Bank,  
  FCA = Financial Conduct Authority (UK),  
  Fed = Federal Reserve (US),  
  FINMA = Swiss Financial Market Supervisory Authority,  
  FSA = Financial Services Agency (Japan),  
  FSC = Financial Services Commission (Korea),  
  HKD = Hong Kong Dollar,  
  HKMA = Hong Kong Monetary Authority,  
  MAS = Monetary Authority of Singapore,  
  OCC = Office of the Comptroller of the Currency (US),  
  PBoC = People’s Bank of China,  
  PSP = Payment Service Provider,  
  SEC = Securities and Exchange Commission (US),  
  SGD = Singapore Dollar.\\[1.2ex]

  \textbf{Interpretive Note:} This table summarizes the evolving regulatory environment for fiat-backed stablecoins across key jurisdictions. Countries like Switzerland and the U.S. foster compliant innovation, while others such as China and Russia prohibit private stablecoins outright. Moderate approaches dominate, balancing innovation with financial stability and AML compliance.

  \end{minipage}
  }
\end{table}

\footnotetext[1]{Effective date not applicable: Private stablecoins are prohibited; only the Digital RMB is legal tender.}
\footnotetext[2]{Effective date not applicable: Russian law bans private digital currencies in favor of the digital ruble.}
\end{landscape}

%% file: tables/llama.tex
\begin{table}[htbp]
\centering
\caption{Key Metrics and Sample Observation from \texttt{yields.llama.fi/pools} (August 3, 2025)}
\label{tab:pool_dictionary}
\scriptsize
\renewcommand{\arraystretch}{1.2}
\rowcolors{2}{gray!5}{white}
\begin{tabular}{@{}p{2.6cm}p{3.2cm}p{5.2cm}@{}}
\toprule
\textbf{Variable} & \textbf{Description} & \textbf{Sample Value (Lido stETH Pool)} \\
\midrule
\multicolumn{3}{l}{\textbf{\textcolor{blue}{Identity and Composition}}} \\
\texttt{project} & Protocol name & \texttt{lido} \\
\texttt{symbol} & Token symbol & \texttt{STETH} \\
\texttt{chain} & Blockchain network & \texttt{Ethereum} \\
\texttt{pool} & Unique pool identifier & \texttt{747c1d2a-c668-4682-b9f9-296708a3dd90} \\
\texttt{underlyingTokens} & List of ERC-20/ETH contracts & \texttt{0x000...000} (ETH) \\
\texttt{stablecoin} & Is it a stablecoin? & \texttt{false} \\
\texttt{exposure} & Exposure type & \texttt{single} \\
\texttt{ilRisk} & Impermanent loss risk & \texttt{no} \\

\midrule
\multicolumn{3}{l}{\textbf{\textcolor{green!50!black}{Yield Performance Metrics}}} \\
\texttt{tvlUsd} & Total value locked (USD) & \$30.80B \\
\texttt{apy} & Net annual percentage yield & 2.632\% \\
\texttt{apyBase} & Yield excluding incentives & 2.632\% \\
\texttt{apyReward} & Bonus/incentive rewards & \texttt{null} \\
\texttt{apyPct1D} & 1-day change in APY & $-0.056$\% \\
\texttt{apyPct7D} & 7-day change in APY & $-0.027$\% \\
\texttt{apyPct30D} & 30-day change in APY & $-0.07$\% \\
\texttt{apyMean30d} & Average APY (30d) & 2.716\% \\
\texttt{apyBaseInception} & Base APY since launch & \texttt{null} \\

\midrule
\multicolumn{3}{l}{\textbf{\textcolor{purple}{Statistical Metrics and Predictions}}} \\
\texttt{mu} & Mean historical APY & 3.760\% \\
\texttt{sigma} & APY standard deviation & 0.052 \\
\texttt{count} & Number of historical samples & 1,157 \\
\texttt{outlier} & Is this an outlier? & \texttt{false} \\
\texttt{predictedClass} & Forecasted APY trend & \texttt{Stable/Up} \\
\texttt{predictedProbability} & Forecast confidence & 74\% \\
\texttt{binnedConfidence} & Confidence bucket & 2 \\
\bottomrule
\end{tabular}
\end{table}